\documentclass[12pt]{article}
\usepackage{amsmath,amssymb}
\usepackage{bm}
\usepackage{color}
\usepackage{cite}
\usepackage{mathtools}

\usepackage{multirow}

\usepackage{here}

\begin{document}

\title{
\begin{flushright}
\ \\*[-80pt]
\begin{minipage}{0.2\linewidth}
\normalsize
EPHOU-22-012\\*[50pt]
\end{minipage}
\end{flushright}
{\Large \bf
Texture zeros of quark mass matrices \\ at fixed point $\tau=\omega$ in modular flavor symmetry
\\*[20pt]}}

\author{
Shota Kikuchi  $^{1}$,
~Tatsuo Kobayashi  $^{1}$, 
\\ Morimitsu Tanimoto $^{2}$,   and~Hikaru Uchida$^{1}$
\\*[20pt]
\centerline{
\begin{minipage}{\linewidth}
\begin{center}
$^1${\it \normalsize
Department of Physics, Hokkaido University, Sapporo 060-0810, Japan} \\*[5pt]
				$^2${\it \normalsize
Department of Physics, Niigata University, Niigata 950-2181, Japan} \\*[5pt]
\end{center}
\end{minipage}}
\\*[50pt]}

\date{
\centerline{\small \bf Abstract}
\begin{minipage}{0.9\linewidth}
\medskip
\medskip
\small
We study systematically derivation of the specific texture zeros, that is the nearest neighbor interaction (NNI) form of the quark mass matrices at the fixed point  $\tau=\omega$ in modular flavor symmetric models.
We present models that the NNI forms of the quark mass matrices are simply realized at the fixed point $\tau=\omega$ in the $A_4$ modular flavor symmetry by taking account multi-Higgs fields.
Such texture zero structure originates from the $ST$ charge of the residual symmetry $Z_3$ of $SL(2,Z)$.
The NNI form can be realized at the fixed point $\tau = \omega$ in $A_4$ and $S_4$ modular flavor models with two pairs of Higgs fields when we assign properly modular weights to Yukawa couplings and $A_4$ and $S_4$ representations to three generations of quarks.
We need four pairs of Higgs fields to realize the NNI form in $A_5$ modular flavor models.
\end{minipage}
}

\begin{titlepage}
\maketitle
\thispagestyle{empty}
\end{titlepage}

\newpage


\section{Introduction}
\label{Intro}

In order to  understand the flavor mixing and the CP violation of the quark and lepton sectors, many works were made to find Ansatz for fermion mass matrices and discussed its predictions. 
The Fritzsch Ansatz \cite{Fritzsch:1977vd,Fritzsch:1979zq} 
was a typical  example. This approach leads to the texture zero analysis where some elements of mass matrices are required to be zero to reduce the degrees of freedom in mass matrices. 
Some famous works have been made in the texture zeros
\cite{Georgi:1979df,Dimopoulos:1991za,Ramond:1993kv}.

Along with those works, the nearest neighbor interaction (NNI) form
\footnote{The NNI form of three families  has vanishing entries of (1,1),\, (2,2),\, (1,3),\, (3,1), but
is  not  necessary to be Hermitian.}
  is considered as a “general” form of both
up- and down-types quark mass matrices because this form is achieved by the transformation that leaves the left- handed gauge interaction invariant
\cite{Branco:1988iq}. 
Based on the NNI  form, some works appeared to explain the flavor mixing of quarks and leptons
\cite{Harayama:1996jr,Harayama:1996am,Takasugi:1997qv,Ito:1996zt,Ito:1997ke}.
The NNI form is a desirable base  to derive the Fritzsch-type quark mass matrix
while the NNI form   is  a general form of quark mass matrices.
 Therefore,  it is  important to study  quark models to realize the NNI form explicitly.


In the recent developments of the modular invariant theories of flavors, the quark and lepton mass matrices are written in terms of modular forms, which are holomorphic functions of the modulus $\tau$ \cite{Feruglio:2017spp}.
Indeed, the well-known finite groups $S_3$, $A_4$, $S_4$ and $A_5$
are isomorphic to the finite modular groups 
$\Gamma_N$ for $N=2,3,4,5$, respectively\cite{deAdelhartToorop:2011re}.
The lepton mass matrices have been given successfully  in terms of {$\Gamma_3\simeq A_4$} modular forms \cite{Feruglio:2017spp}.
Modular invariant flavor models have also been proposed on the $\Gamma_2\simeq  S_3$ \cite{Kobayashi:2018vbk},
$\Gamma_4 \simeq  S_4$ \cite{Penedo:2018nmg} and  
$\Gamma_5 \simeq  A_5$ \cite{Novichkov:2018nkm,Ding:2019xna}.
Other finite groups are also derived from magnetized D-brane models \cite{Kobayashi:2018bff}.
By using these modular forms, the flavor mixing of quarks and leptons has been discussed successfully in these years
since the non-Abelian finite groups are long familiar
in  quarks and leptons
\cite{Altarelli:2010gt,Ishimori:2010au,Ishimori:2012zz,Kobayashi:2022moq,Hernandez:2012ra,King:2013eh,King:2014nza,Tanimoto:2015nfa,King:2017guk,Petcov:2017ggy,Feruglio:2019ktm}.

Phenomenological studies of the lepton flavors have been done
based on  $A_4$ \cite{Criado:2018thu,Kobayashi:2018scp,Ding:2019zxk}, 
$S_4$ \cite{Novichkov:2018ovf,Kobayashi:2019mna,Wang:2019ovr} and 
$ A_5$ \cite{Novichkov:2018nkm,Ding:2019xna}.
Furthermore, phenomenological studies have been developed  in many works.
Among them,  the texture zeros of quarks and leptons  have been discussed
 in the context of the assignment of the weight for the chiral superfields \cite{Zhang:2019ngf,Lu:2019vgm}.
 
 However, the realization of texture zeros is not necessary to adjust the weight of the chiral superfields.
For example,  the fermion mass matrix has the texture zero structure at $\tau=\omega=e^{2 \pi i/3}$ due to the $Z_3$ symmetry  independent of the weights \cite{Okada:2020ukr,Kobayashi:2021pav} although the flavor mixing are not reproduced.
That is, the theory becomes special due to residual symmetries at the fixed points of $SL(2,Z)$
\cite{Feruglio:2021dte,Novichkov:2021evw}.
The fixed point $\tau=\omega$ is also favored from the moduli stabilization.
This fixed point has the highest probability in the moduli stabilization due to 
three-form background \cite{Ishiguro:2020tmo}.
Other moduli stabilization mechanisms were studied in 
modular flavor models \cite{Kobayashi:2019uyt,Abe:2020vmv,Novichkov:2022wvg,Ishiguro:2022pde}.
The fixed points are also useful to stabilize dark matter candidates \cite{Kobayashi:2021ajl}.

Recently,  the CP violation   at $\tau=\omega$ has been discussed in
 magnetized orbifold models with multi-Higgs modes \cite{Kikuchi:2022geu}.
Magnetized orbifold models are interesting compactification from higher dimensional theory such as superstring theory.
They lead to a four-dimensional chiral theory, where the generation number is determined by 
the size of magnetic flux in the compact space \cite{Abe:2008fi,Abe:2013bca}.
The four-dimensional low-energy effective field theory has the modular symmetry \cite{Kobayashi:2018rad,Kobayashi:2018bff,Ohki:2020bpo,Kikuchi:2020frp,Kikuchi:2020nxn,
Kikuchi:2021ogn,Almumin:2021fbk}.
Realization of quark and lepton masses and their mixing angles was studied  \cite{Abe:2012fj,Abe:2014vza,Fujimoto:2016zjs}.
Their texture structures were also studied \cite{Kikuchi:2021yog}.
These magnetized orbifold models lead to multi-Higgs modes, while generic string compactification also 
leads more than one candidates for Higgs fields.

 In this work,  we present models that the NNI forms of the quark mass matrices are simply realized  at the fixed point 
 $\tau=\omega$ of the modular symmetry  by taking account  multi-Higgs fields.
We use the $A_4$ modular symmetry as well as $S_4$ and $A_5$.
The quark mass matrices with texture zeros, which are consistent with observed CKM matrix elements,
  are also derived.
 These models are also simple examples that the CP is violated even at $\tau=\omega$.

The paper is organized as follows.
In section 2, we present a simple example of  the NNI form at $\tau=\omega$ in $A_4$ modular symmetry.
In section 3, we study a generic model systematically.
Section 4 is our conclusion.
We summarize group theoretical aspects of $A_4$, $S_4$, and $A_5$ in Appendix A and
  the modular forms of level $N=3$ in Appendix B. 

\section{NNI form of quark mass matrices at $\tau=\omega$}
\subsection{Quark mass matrices with mluti-Higgs}
In this section,
  we present a simple model of quark mass matrices 
   in the level $N=3$ modular symmetry ($A_4$ modular flavor symmetry)
    with the multi-Higgs at $\tau=\omega$, which is referred  to as Model 1.
We assign  the $A_4$ representation and the weights for the relevant chiral superfields as
\begin{itemize}
  \item{quark doublet $Q=(Q^1,Q^2,Q^3)$: $A_4$ triplet with weight -2}
  \item{up-type quark singlets $(u,c,t)$: $A_4$ singlets $(1,1',1'')$ with weight 0}
  \item{down-type quark singlets $(d,s,b)$: $A_4$ singlets $(1,1',1'')$ with weight 0}
  \item{up and down sector Higgs fields $H_{u,d}^i=(H_{u,d}^1,H_{u,d}^2)$: $A_4$ singlets $(1,1'')$ with weight 0}
\end{itemize}
which are summarized in Table \ref{tab:1}.
\begin{table}[h]
\begin{center}
\renewcommand{\arraystretch}{1.1}
\begin{tabular}{|c|c|c|c|c|c|} \hline
  & $Q=(Q^1,Q^2,Q^3)$ & $(u,c,t)$ & $(d,s,b)$ & $H_u$ & $H_d$ \\ \hline
  $SU(2)$ & 2 & 1 & 1 & 2 & 2 \\
  $A_4$ & 3 & $(1,1',1'')$ & $(1,1',1'')$ & $(1,1'')$ & $(1,1'')$ \\
  $k$ & -2 & 0 & 0 & 0 & 0 \\ \hline
\end{tabular}
\end{center}
\caption{Assignments of $A_4$ representations and weights in Model 1.}
\label{tab:1}
\end{table}

Then, the superpotential terms of the up-type quark masses and down-type quark masses are written by
\begin{align}
  &W_u =
  \left[\alpha^{1}_u ({\bf Y^{(2)}}Q)_1 u_{1} + \beta^{1}_u ({\bf Y^{(2)}}Q)_{1''}c_{1'} + \gamma^{1}_u ({\bf Y^{(2)}}Q)_{1'}t_{1''}\right](H_u^1)_1  \nonumber\\
  &\quad + \left[\alpha^{2}_u ({\bf Y^{(2)}}Q)_{1'} u_{1} + \beta^{2}_u ({\bf Y^{(2)}}Q)_{1}c_{1'} + \gamma^{2}_u ({\bf Y^{(2)}}Q)_{1''}t_{1''}\right](H_u^2)_{1''}, \\
  \nonumber\\\
  &W_d =
  \left[\alpha^{1}_d ({\bf Y^{(2)}}Q)_1 d_{1} + \beta^{1}_d ({\bf Y^{(2)}}Q)_{1''}s_{1'} + \gamma^{1}_d ({\bf Y^{(2)}}Q)_{1'}b_{1''}\right](H_d^1)_1 \nonumber\\\
  &\quad + \left[\alpha^{2}_d ({\bf Y^{(2)}}Q)_{1'} d_{1} + \beta^{2}_d ({\bf Y^{(2)}}Q)_{1}s_{1'} + \gamma^{2}_d ({\bf Y^{(2)}}Q)_{1''}b_{1''}\right](H_d^2)_{1''},
\end{align}
where the decompositions of the tensor products are
\begin{align}
  &({\bf Y^{(2)}}Q)_1 = 
  \left(
  \begin{pmatrix}
    Y_1\\Y_2\\Y_3\\
  \end{pmatrix}_3
  \otimes
  \begin{pmatrix}
    Q^1\\Q^2\\Q^3\\
  \end{pmatrix}_3
  \right)_1
  = Y_1Q^1+Y_2Q^3+Y_3Q^2, \\
  &({\bf Y^{(2)}}Q)_{1''} = 
  \left(
  \begin{pmatrix}
    Y_1\\Y_2\\Y_3\\
  \end{pmatrix}_3
  \otimes
  \begin{pmatrix}
    Q^1\\Q^2\\Q^3\\
  \end{pmatrix}_3
  \right)_{1''}
  = Y_3Q^3+Y_1Q^2+Y_2Q^1, \\
  &({\bf Y^{(2)}}Q)_{1'} = 
  \left(
  \begin{pmatrix}
    Y_1\\Y_2\\Y_3\\
  \end{pmatrix}_3
  \otimes
  \begin{pmatrix}
    Q^1\\Q^2\\Q^3\\
  \end{pmatrix}_3
  \right)_{1'}
  = Y_2Q^2+Y_1Q^3+Y_3Q^1.
\end{align}

The superpotential terms are rewritten as:
\begin{align}
  W_u &= [\alpha_u^1(Y_1Q^1+Y_2Q^3+Y_3Q^2)u+\beta^1_u(Y_3Q^3+Y_1Q^2+Y_2Q^1)c+\gamma^1_u(Y_2Q^2+Y_1Q^3+Y_3Q^1)t]H_u^1 \notag \\
  &+[\alpha_u^2(Y_2Q^2+Y_1Q^3+Y_3Q^1)u+\beta^2_u(Y_1Q^1+Y_2Q^3+Y_3Q^2)c+\gamma^2_u(Y_3Q^3+Y_1Q^2+Y_2Q^1)t]H_u^2 \notag \\
  &=
  \begin{pmatrix}
    Q^1&Q^2&Q^3\\
  \end{pmatrix}
  \left(
  \begin{pmatrix}
    \alpha_u^1Y_1 & \beta_u^1Y_2 & \gamma_u^1Y_3 \\
    \alpha_u^1Y_3 & \beta_u^1Y_1 & \gamma_u^1Y_2 \\
    \alpha_u^1Y_2 & \beta_u^1Y_3 & \gamma_u^1Y_1 \\
  \end{pmatrix}H_u^1
  +
  \begin{pmatrix}
    \alpha_u^2Y_3 & \beta_u^2Y_1 & \gamma_u^2Y_2 \\
    \alpha_u^2Y_2 & \beta_u^2Y_3 & \gamma_u^2Y_1 \\
    \alpha_u^2Y_1 & \beta_u^2Y_2 & \gamma_u^2Y_3 \\
  \end{pmatrix}H_u^2
  \right)
  \begin{pmatrix}
    u\\c\\t\\
  \end{pmatrix}, \\
  W_d &= [\alpha_d^1(Y_1Q^1+Y_2Q^3+Y_3Q^2)d+\beta^1_d(Y_3Q^3+Y_1Q^2+Y_2Q^1)s+\gamma^1_d(Y_2Q^2+Y_1Q^3+Y_3Q^1)b]H_d^1 \notag \\
  &+[\alpha_d^2(Y_2Q^2+Y_1Q^3+Y_3Q^1)d+\beta^2_d(Y_1Q^1+Y_2Q^3+Y_3Q^2)s+\gamma^2_d(Y_3Q^3+Y_1Q^2+Y_2Q^1)b]H_d^2 \notag \\
  &=
  \begin{pmatrix}
    Q^1&Q^2&Q^3\\
  \end{pmatrix}
  \left(
  \begin{pmatrix}
    \alpha_d^1Y_1 & \beta_d^1Y_2 & \gamma_d^1Y_3 \\
    \alpha_d^1Y_3 & \beta_d^1Y_1 & \gamma_d^1Y_2 \\
    \alpha_d^1Y_2 & \beta_d^1Y_3 & \gamma_d^1Y_1 \\
  \end{pmatrix}H_d^1
  +
  \begin{pmatrix}
    \alpha_d^2Y_3 & \beta_d^2Y_1 & \gamma_d^2Y_2 \\
    \alpha_d^2Y_2 & \beta_d^2Y_3 & \gamma_d^2Y_1 \\
    \alpha_d^2Y_1 & \beta_d^2Y_2 & \gamma_d^2Y_3 \\
  \end{pmatrix}H_d^2
  \right)
  \begin{pmatrix}
    d\\s\\b\\
  \end{pmatrix}.
\end{align}

Finally, the quark mass matrices  are  given as:
\begin{align}
  M_u &=
  \begin{pmatrix}
    \alpha_u^1Y_1 & \beta_u^1Y_2 & \gamma_u^1Y_3 \\
    \alpha_u^1Y_3 & \beta_u^1Y_1 & \gamma_u^1Y_2 \\
    \alpha_u^1Y_2 & \beta_u^1Y_3 & \gamma_u^1Y_1 \\
  \end{pmatrix} \langle H_u^1\rangle
  +
  \begin{pmatrix}
    \alpha_u^2Y_3 & \beta_u^2Y_1 & \gamma_u^2Y_2 \\
    \alpha_u^2Y_2 & \beta_u^2Y_3 & \gamma_u^2Y_1 \\
    \alpha_u^2Y_1 & \beta_u^2Y_2 & \gamma_u^2Y_3 \\
  \end{pmatrix} \langle H_u^2\rangle, \\
  M_d &=
  \begin{pmatrix}
    \alpha_d^1Y_1 & \beta_d^1Y_2 & \gamma_d^1Y_3 \\
    \alpha_d^1Y_3 & \beta_d^1Y_1 & \gamma_d^1Y_2 \\
    \alpha_d^1Y_2 & \beta_d^1Y_3 & \gamma_d^1Y_1 \\
  \end{pmatrix} \langle H_u^1\rangle
  +
  \begin{pmatrix}
    \alpha_d^2Y_3 & \beta_d^2Y_1 & \gamma_d^2Y_2 \\
    \alpha_d^2Y_2 & \beta_d^2Y_3 & \gamma_d^2Y_1 \\
    \alpha_d^2Y_1 & \beta_d^2Y_2 & \gamma_d^2Y_3 \\
  \end{pmatrix} \langle H_u^2\rangle,
\end{align}
where the chiralities of the mass matrix, $L$ and  $R$ are defined as $[M_{u(d)}]_{LR}$.

\subsection{$ST$-eigenstate base at $\tau=\omega$}
Let us discuss the mass matrices at $\tau=\omega$ in the $ST$-eigenstates.
The $ST$-transformation of the $A_4$ triplet of the left-handed quarks $Q$ is
\begin{align}
  \begin{pmatrix}
    Q^1 \\ Q^2 \\ Q^3\\
  \end{pmatrix}
  &\xrightarrow{ST}
  (-\omega-1)^{-2}
  \rho(ST) 
  \begin{pmatrix}
    Q^1 \\ Q^2 \\ Q^3\\
  \end{pmatrix}   \nonumber\\\
  &= \omega^{-4}
  \frac{1}{3}
  \begin{pmatrix}
    -1 & 2\omega & 2\omega^2 \\
    2 & -\omega & 2\omega^2 \\
    2 & 2\omega & -\omega^2 \\
  \end{pmatrix}
  \begin{pmatrix}
    Q^1 \\ Q^2 \\ Q^3\\
  \end{pmatrix},
\end{align}
where representations of $S$ and $T$ are given explicitly for the triplet in Appendix A.
The $ST$-eigenstate $Q'$ is obtained by using the unitary matrix $U_L$ as follows:
\begin{align}
  &U_L = \frac{1}{3}
  \begin{pmatrix}
    2 & -\omega & 2\omega^2 \\
    -\omega & 2\omega^2 & 2 \\
    2\omega^2 & 2 & -\omega \\
  \end{pmatrix}, \\
  &U_L^\dagger \omega^{-4} \rho(ST) U_L =
  \begin{pmatrix}
    1 & 0 & 0 \\
    0 & \omega^2 & 0 \\
    0 & 0 & \omega \\
  \end{pmatrix}.
\end{align}
The $ST$-eigenstates are
$Q'\equiv U_L^\dagger Q$.

On the other hand,
right-handed quarks, which are singlets $(1,1',1'')$, are the eigenstates of $ST$;
that is, 
the $ST$-transformation is
\begin{align}
  &\begin{pmatrix}
    u \\ c \\ t \\
  \end{pmatrix}
  \xrightarrow{ST}
  \begin{pmatrix}
    1 & 0 & 0 \\
    0 & \omega^2 & 0 \\
    0 & 0 & \omega \\
  \end{pmatrix}
  \begin{pmatrix}
    u \\ c \\ t \\
  \end{pmatrix}, \quad
  \begin{pmatrix}
    d \\ s \\ b \\
  \end{pmatrix}
  \xrightarrow{ST}
  \begin{pmatrix}
    1 & 0 & 0 \\
    0 & \omega^2 & 0 \\
    0 & 0 & \omega \\
  \end{pmatrix}
  \begin{pmatrix}
    d \\ s \\ b \\
  \end{pmatrix}.
\end{align}

Higgs fields are also the $ST$-eigenstates since they are singlets $(1,1'')$.
Therefore, $ST$-transformation of them is
\begin{align}
  &\begin{pmatrix}
    H_{u,d}^1 \\ H_{u,d}^2 \\
  \end{pmatrix}
  \xrightarrow{ST}
  \begin{pmatrix}
    1 & 0 \\
    0 & \omega \\
  \end{pmatrix}
  \begin{pmatrix}
    H_{u,d}^1 \\ H_{u,d}^2 \\
  \end{pmatrix}.
\end{align}

In the $ST$-eigenstates,  the quark mass matrices are given as:
\begin{align}
  &U_L^T M_u = c
  \begin{pmatrix}
    \alpha_u^1 & 0 & 0 \\
    0 & 0 & \gamma_u^1 \\
    0 & \beta_u^1 & 0 \\
  \end{pmatrix}\langle H_u^1\rangle
  + c
  \begin{pmatrix}
    0 & \beta_u^2 & 0 \\
    \alpha_u^2 & 0 & 0 \\
    0 & 0 & \gamma_u^2 \\
  \end{pmatrix}\langle H_u^2\rangle, 
  \label{quark-mass-matrix-up}\\
  &U_L^T M_d = c
  \begin{pmatrix}
    \alpha_d^1 & 0 & 0 \\
    0 & 0 & \gamma_d^1 \\
    0 & \beta_d^1 & 0 \\
  \end{pmatrix}\langle H_d^1\rangle
  + c
  \begin{pmatrix}
    0 & \beta_d^2 & 0 \\
    \alpha_d^2 & 0 & 0 \\
    0 & 0 & \gamma_d^2 \\
  \end{pmatrix}\langle H_d^2\rangle,
  \label{quark-mass-matrix-down}
\end{align}
where $c=\sqrt{|Y_1|^2+|Y_2|^2+|Y_3|^2}$.
Their ratios at $\tau =\omega$ are obtained as
\begin{align}
(Y_1(\omega),Y_2(\omega),Y_3(\omega))=Y_1(\omega)(1,\omega,-\frac12 \omega^2).
\end{align}
Now,  imposing  $\alpha^1_{u,d}=0$, we obtain the NNI forms  for both the up-type and the down-type quark mass matrices.
Therefore, the quark masses and the CKM matrix are reproduced taking relevant values
of parameters.
It is noticed that the flavor mixing is not realized in the case of one Higgs doublets for up- and down-type quark sectors. 
Thus, the NNI forms at $\tau=\omega$  are simply obtained 
unless the vacuum expectation values (VEVs) of two-Higgs vanish.
The general discussion is presented  in section~\ref{sec:NNI}.

The CP symmetry is not violated at $\tau = \omega$ in modular flavor symmetric models with 
a pair of Higgs fields because of the $T$ symmetry \cite{Kobayashi:2019uyt}.
However, the models with multi-Higgs fields can break the CP symmetry at 
the fixed point $\tau = \omega$ even if all of the Higgs VEVs are real \cite{Kikuchi:2022geu}.
Thus,  the CP phase appears in our  models, in general.
Our models are interesting from the viewpoint of the CP violation, too.

The non-vanishing VEVs of both Higgs fields $H^1_{u,d}$ and $H^2_{u,d}$ are important to 
realize the NNI forms.
We expect the scenario that these Higgs fields have a $\mu$-matrix to 
mix them,
\begin{align}
W_\mu = \mu_{ij}H^i_u H^j_d .
\end{align}
Then, a light linear combination develops its VEV, which includes 
$H^1_{u,d}$ and $H^2_{u,d}$.
However, the above assignment of modular weights for the Higgs fields allows 
the $\mu$-term of only $\mu_{11}$, and the others vanish.
That is, the mixing does not occur.
When we assume the singlet $S$ with the $A_4$ $1'$ representation develops its VEV, 
the $(1,2)$ and $(2,1)$ elements appear as $\mu_{12}=\mu_{21}=\lambda \langle S \rangle$ 
like the next-to-minimal supersymmetric standard model.

Alternatively, we can assign the modular weights to fields as shown in Table \ref{tab:model2}.
We refer to this model as Model 2.
By this assignment, we can obtain the same quark mass matrices as 
one in Eqs.\eqref{quark-mass-matrix-up} and \eqref{quark-mass-matrix-down}.
The superpotential terms for  Higgs $\mu$-terms are given in terms of the weight 4 modular forms as:
\begin{align}
  W_H &= g_{11}({\bf Y^{(4)}}H_u^1H_d^1)_1+g_{12}({\bf Y^{(4)}}H_u^1H_d^2)_1 + g_{21}({\bf Y^{(4)}}H_u^2H_d^1)_1 + g_{22}({\bf Y^{(4)}}H_u^2H_d^2)_1 \nonumber\\
   \nonumber\\
  &= g_{11}({\bf Y^{(4)}})_1H_u^1H_d^1+g_{12}({\bf Y^{(4)}})_{1'}H_u^1H_d^2 + g_{21}({\bf Y^{(4)}})_{1'}H_u^2H_d^1 + g_{22}({\bf Y^{(4)}})_{1''}H_u^2H_d^2,
\end{align}
where
\begin{align}
  {\bf Y^{(4)}} = ({\bf Y}^{\bf (4)}_1, {\bf Y}^{\bf (4)}_{1'}, {\bf Y}^{\bf (4)}_3)
\end{align}
are given in Appendix B.
\begin{table}[h]
\begin{center}
\renewcommand{\arraystretch}{1.1}
\begin{tabular}{|c|c|c|c|c|c|} \hline
  & $Q=(Q^1,Q^2,Q^3)$ & $(u,c,t)$ & $(d,s,b)$ & $H_u$ & $H_d$ \\ \hline
  $SU(2)$ & 2 & 1 & 1 & 2 & 2 \\
  $A_4$ & 3 & $(1,1',1'')$ & $(1,1',1'')$ & $(1,1'')$ & $(1,1'')$ \\
  $k$ & 0 & 0 & 0 & -2 & -2 \\ \hline
\end{tabular}
\end{center}
\caption{Assignments of Model 2}
\label{tab:model2}
\end{table}

The superpotential $W_H$ is explicitly  given as:
\begin{align}
  W_H &=
  \begin{pmatrix}
    H_u^1 & H_u^2 \\
  \end{pmatrix}
  \begin{pmatrix}
    g_{11}({\bf Y^{(4)}})_1 & g_{12}({\bf Y^{(4)}})_{1'} \\
    g_{21}({\bf Y^{(4)}})_{1'} & g_{22}({\bf Y^{(4)}})_{1''} \\
  \end{pmatrix}
  \begin{pmatrix}
    H_d^1 \\ H_d^2
  \end{pmatrix}   \nonumber\\
  &=
  \begin{pmatrix}
    H_u^1 & H_u^2  \\
  \end{pmatrix}
  \begin{pmatrix}
    g_{11}(Y_1^2+2Y_2Y_3) & g_{12}(Y_3^2+2Y_1Y_2) \\
    g_{21}(Y_3^2+2Y_1Y_2) & 0 \\
  \end{pmatrix}
  \begin{pmatrix}
    H_d^1 \\ H_d^2
  \end{pmatrix}.
\end{align}
Thus, the Higgs fields $H^1_{u,d}$ and $H^2_{u,d}$ mix in their mass spectrum.
When we assume that the lightest mode develops its VEV, 
we have non-vanishing VEVs of $\langle H^{1}_{u,d} \rangle$ and 
$\langle H^2_{u,d} \rangle$.
We denote 
\begin{align}
\mu_{11} = g_{11}(Y_1^2+2Y_2Y_3), \qquad \mu_{12} =g_{12}(Y_3^2+2Y_1Y_2).
\end{align}
Then, mass eigenvalues  are written by
\begin{align}
m_{\pm} = \frac12 \left(\mu_{11} \pm \sqrt{\mu_{11}^2 + 4\mu_{12}^2} \right),
\end{align}
and the corresponding Higgs directions are written by 
\begin{align}
m_{\pm}H^1_{u,d}+\mu_2 H^2_{u,d},
\end{align}
up to normalization factors.
For example, when $\mu_{11} \gg \mu_{12}$, we obtain 
\begin{align}
m_+ \approx \mu_{11}, \qquad m_- \approx  \frac{\mu_{12}^2}{\mu_{11}}.
\end{align}
Then, the light mode corresponds to the following direction:
\begin{align}
\frac{\mu_{12}}{\mu_{11}}H^1_{u,d} + H^2_{u,d},
\end{align}
up to a normalization factor.

Heavier Higgs modes would contribute to flavor changing processes.
They depend on the mass of the heavier modes, which are free parameters 
in the above model.
Those flavor changing processes are suppressed when 
heavier modes are heavy enough.
Studies on flavor changing processes would be important 
if we have a scenario to predict the mass scale of heavier modes.
That is beyond our scope.


\subsection{Three pairs of  Higgs fields $(1,1'',1')$}

Similarly, we can study three pairs of Higgs fields with the $A_4$ $(1,1'',1')$ representations.
We add another pair  of Higgs fields $H^3_{u,d}$ with the $A_4$ $1'$ representation of the modular weights, 
0 and $-2$ in models 1 and 2, respectively.
Then, the mass matrices are modified as follows:
\begin{align}
  &U_L^T M_u = c
  \begin{pmatrix}
    \alpha_u^1 & 0 & 0 \\
    0 & 0 & \gamma_u^1 \\
    0 & \beta_u^1 & 0 \\
  \end{pmatrix}\langle H_u^1\rangle
  + c
  \begin{pmatrix}
    0 & \beta_u^2 & 0 \\
    \alpha_u^2 & 0 & 0 \\
    0 & 0 & \gamma_u^2 \\
  \end{pmatrix}\langle H_u^2\rangle
  + c
  \begin{pmatrix}
    0 & 0 & \gamma_u^3 \\
    0 & \beta_u^3 & 0 \\
    \alpha_u^3 & 0 & 0 \\
  \end{pmatrix}\langle H_u^3\rangle, \\
  &U_L^T M_d = c
  \begin{pmatrix}
    \alpha_d^1 & 0 & 0 \\
    0 & 0 & \gamma_d^1 \\
    0 & \beta_d^1 & 0 \\
  \end{pmatrix}\langle H_d^1\rangle
  + c
  \begin{pmatrix}
    0 & \beta_d^2 & 0 \\
    \alpha_d^2 & 0 & 0 \\
    0 & 0 & \gamma_d^2 \\
  \end{pmatrix}\langle H_d^2\rangle
  + c
  \begin{pmatrix}
    0 & 0 & \gamma_d^3 \\
    0 & \beta_d^3 & 0 \\
    \alpha_d^3 & 0 & 0 \\
  \end{pmatrix}\langle H_d^3\rangle, 
\end{align}
in both models 1 and 2.
 Thus, this model can lead to a quite generic mass matrix.
For example, by setting some of $\alpha^i_{u,d}, \beta^i_{u,d}, \gamma^i_{u,d}$ to be zero, 
we can drive some of texture zero structures including the NNI form.
In addition, we can assume $\beta^i_{u,d}=\gamma^i_{u,d}$ or $\beta^i_{u,d}=(\gamma^i_{u,d})^*$ 
to reduce the number of free parameters and realize a certain form of mass matrices.
Thus, the different assignment of the $A_4$ singlets $(1,1'',1')$  for Higgs leads to different 
 texture zeros.



\section{Generic models}
\label{sec:NNI}

 In the previous section,  the quark mass matrices are discussed in the specific modular symmetry
 of $N=3$ in order to show the derivation of NNI forms clearly.
Similarly, we can study a generic mode leading to the NNI forms 
including $S_4$ and $A_5$ modular flavor symmetries.

\subsection{Residual $Z_3$ symmetry}

Because of $(ST)^3=1$, each field under the $ST$ basis has the 
$Z_3^{(ST)}$ charge.
Here, let us  discuss the $ST$ charge assignment without specifying the finite modular groups.
At first, consider the case of the single Higgs field of up-type quark sector for simplicity 
as seen in Table \ref{tab:ST-charge}.
Quarks may belong to a multiplet such as a triplet, 
but we study just the $ST$ charge.

\begin{table}[H]
\begin{center}
\renewcommand{\arraystretch}{1.1}
\begin{tabular}{c|c|c|c} \hline
  & $Q=(Q^1,Q^2,Q^3)$ & $(u,c,t)$ & $H_u$ \\ \hline
  $Z_3^{(ST)}$ charge & $(0,1,2)$ & $(0,1,2)$ & $0$ \\ \hline
\end{tabular}
\end{center}
\caption{$ST$ charges of fields of up-type quark sector}
\label{tab:ST-charge}
\end{table}
\noindent

The modular forms transform as:
\begin{align}
  f^i(\tau) \rightarrow (-\tau-1)^{k}\rho^{ij}(ST) f^j(\tau) 
   \end{align}
where the $Z_3^{(ST)}$ charge $q_i$ is defined as
\begin{align}
(e^{2 \pi i q_1/3}, e^{2 \pi i q_2/3},e^{2 \pi i q_3/3})=
{\rm diag}((-\tau-1)^k\rho^{ij}(ST)) \, ,
\end{align}
including the automorphic factor.
Then, the superpotential for up-type quark mass matrix is given as
\begin{align}
  W(\tau) = \sum_i \alpha^i_u (Y_{{\bf r}_i}(\tau)Qu H_u)_1 + \sum_i \alpha^i_c (Y_{{\bf r}_i}(\tau)Qc H_u)_1 + \sum_i \alpha^i_t (Y_{{\bf r}_i}(\tau)Qt H_u)_1
\end{align}
where coefficients of the singlet components are written as:
\begin{align}
  (Y_{{\bf r}_i}(\tau) Qu H_u)_1 &= C_u^{ijk}Y_{{\bf r}_i}^j(\tau)Q^kuH_u, \\
  (Y_{{\bf r}_i}(\tau) Qc H_u)_1 &= C_c^{ijk}Y_{{\bf r}_i}^j(\tau)Q^kcH_u, \\
  (Y_{{\bf r}_i}(\tau) Qt H_u)_1 &= C_t^{ijk}Y_{{\bf r}_i}^j(\tau)Q^ktH_u.
\end{align}
Under the  $ST$ transformation of $Y_{{\bf r}_i}^j$ is invariant at $\tau=\omega$,
 $Y_{{\bf r}_i}^j$ vanishes unless $Z_3^{(ST)}$ charge is 0, because 
\begin{align}
  Y^j_{{\bf r}_i}(\omega) \rightarrow Y^j_{{\bf r}_i}(ST\omega) = (-\omega-1)^{k_Y}\rho^{jk}(ST) Y^k_{{\bf r}_i}(\omega) = \omega^{2k_Y}\rho^{jk}(ST) Y^k_{{\bf r}_i}(\omega) \,,
\end{align}
where $\omega^{2k_Y}\rho^{jk}(ST)$ corresponds to the $ST$ charge.
Therefore, we obtain non-vanishing components of the mass matrix from the assignment of $ST$ charges for fields in Table 3:
\begin{align}
  {(Y_{{\bf r}_i}(\omega) Qu H_u)_1}  &= C_u^{ij1}Y_{{\bf r}_i}^j(\omega){Q^1uH_u}
\end{align}
\begin{align}
  (Y_{{\bf r}_i}(\omega) Qc H_u)_1 &= C_c^{ij3}Y_{{\bf r}_i}^j(\omega)Q^3cH_u, \\
  \nonumber\\
  (Y_{{\bf r}_i}(\omega) Qt H_u)_1 &= C_t^{ij2}Y_{{\bf r}_i}^j(\omega)Q^2tH_u\,.
\end{align}
Then, the mass matrix is given as:
\begin{align}
  W(\omega) &= \alpha^i_u C_u^{ij1}Y_{{\bf r}_i}^j(\omega) Q^1uH_u + \alpha^i_c C_c^{ij3}Y_{{\bf r}_i}^j(\omega)Q^3cH_u + \alpha^i_t C_t^{ij2}Y_{{\bf r}_i}^j(\omega)Q^2tH_u \\
  &= 
  \begin{pmatrix}
    Q^1 & Q^2 & Q^3 \\
  \end{pmatrix}
  \begin{pmatrix}
    \alpha^i_u C_u^{ij1}Y_{{\bf r}_i}^j(\omega) & 0 & 0 \\
    0 & 0 & \alpha^i_t C_t^{ij2}Y_{{\bf r}_i}^j(\omega) \\
    0 & \alpha^i_c C_c^{ij3}Y_{{\bf r}_i}^j(\omega) & 0 \\
  \end{pmatrix}
  \begin{pmatrix}
    u \\ c \\ t \\
  \end{pmatrix} H_u \,.
\end{align}

If we take $ST$ charge for the Higgs field  is $1$ as in Table \ref{tab:ST-charge-1},
the mass matrix is given as:
\begin{align}
  W(\omega) &= \alpha^i_u C_u^{ij3}Y_{{\bf r}_i}^j(\omega) Q^3uH_u + \alpha^i_c C_c^{ij2}Y_{{\bf r}_i}^j(\omega)Q^2cH_u + \alpha^i_t C_t^{ij1}Y_{{\bf r}_i}^j(\omega)Q^1tH_u \\
  &= 
  \begin{pmatrix}
    Q^1 & Q^2 & Q^3 \\
  \end{pmatrix}
  \begin{pmatrix}
    0 & 0 & \alpha^i_t C_t^{ij1}Y_{{\bf r}_i}^j(\omega) \\
    0 & \alpha^i_c C_c^{ij2}Y_{{\bf r}_i}^j(\omega) & 0 \\
    \alpha^i_u C_u^{ij3}Y_{{\bf r}_i}^j(\omega) & 0 & 0 \\
  \end{pmatrix}
  \begin{pmatrix}
    u \\ c \\ t \\
  \end{pmatrix} H_u.
\end{align}
\begin{table}[H]
\begin{center}
\renewcommand{\arraystretch}{1.1}
\begin{tabular}{c|c|c|c} \hline
  & $Q=(Q^1,Q^2,Q^3)$ & $(u,c,t)$ & $H_u$ \\ \hline
  $Z_3^{(ST)}$ charge & $(0,1,2)$ & $(0,1,2)$ & $1$ \\ \hline
\end{tabular}
\end{center}
\caption{$ST$ charges of fields of up-type quark sector}
\label{tab:ST-charge-1}
\end{table}

Finally, taking $ST$-charge for the Higgs field to be $2$ as in Table \ref{tab:ST-charge-2},
the mass matrix is:
\begin{align}
  W(\omega) &= \alpha^i_u C_u^{ij2}Y_{{\bf r}_i}^j(\omega) Q^2uH_u + \alpha^i_c C_c^{ij1}Y_{{\bf r}_i}^j(\omega)Q^1cH_u + \alpha^i_t C_t^{ij3}Y_{{\bf r}_i}^j(\omega)Q^3tH_u \\
  &= 
  \begin{pmatrix}
    Q^1 & Q^2 & Q^3 \\
  \end{pmatrix}
  \begin{pmatrix}
    0 & \alpha^i_c C_c^{ij1}Y_{{\bf r}_i}^j(\omega) & 0 \\
    \alpha^i_u C_u^{ij2}Y_{{\bf r}_i}^j(\omega) & 0 & 0 \\
    0 & 0 & \alpha^i_t C_t^{ij3}Y_{{\bf r}_i}^j(\omega) \\
  \end{pmatrix}
  \begin{pmatrix}
    u \\ c \\ t \\
  \end{pmatrix} H_u\,.
\end{align}
\begin{table}[H]
\begin{center}
\renewcommand{\arraystretch}{1.1}
\begin{tabular}{c|c|c|c} \hline
  & $Q=(Q^1,Q^2,Q^3)$ & $(u,c,t)$ & $H_u$ \\ \hline
  $Z_3^{(ST)}$ charge & $(0,1,2)$ & $(0,1,2)$ & $2$ \\ \hline
\end{tabular}
\end{center}
\caption{$ST$ charges of fields of up-type quark sector}
\label{tab:ST-charge-2}
\end{table}

By combining these matrices, we can obtain the NNI form as well as other textures.
Thus, the $Z_3^{(ST)}$ symmetry is important to construct the NNI form.
Three generations of quarks should have $Z_3^{(ST)}$ charges different from 
each other.
At the fixed point $\tau =i$, there remains the $Z_2^{(S)}$ symmetry in $PSL(2,Z)$.
For the $Z_2^{(S)}$ symmetry, two generations among three generations of quarks 
must have the same $Z_2^{(S)}$ charge.
Hence, we can not realize the above form, which can be derived by the 
$Z_3^{(ST)}$ symmetry.
On the other hand, at the limit $\tau \to i \infty$, 
the $Z_N^{(T)}$ symmetry remains for $\Gamma_N$, i.e. $T^N=1$.
Such a residual symmetry may be useful to construct the NNI form.
However, the limit $\tau \to i \infty$ corresponds to 
decompactification in extra dimensional theory such as superstring theory.

The above discussion is  generic  without specifying representations.
When we specify the representations, the mass matrices are constrained more.
Table \ref{tab:representations} shows irreducible representations of $A_4$, $S_4$, and $A_5$.
In addition, Tables \ref{tab:A4-form}, \ref{tab:S4-form}, \ref{tab:A5-form} show the relations between representations and weighs of modular forms of 
$A_4$, $S_4$, $A_5$, respectively.
For example, when left-handed and right-handed quarks are assigned to 
the same triplet, the mass matrix must be symmetric.
In addition, the coefficients $\alpha^i_{u,d}$, $\beta^i_{u,d}$, $\gamma^i_{u.d}$ are not 
independent parameters, but must be related.
We do not have a sufficient number of free parameters to 
realize realistic masses and mixing angles.
Thus, it is important how to assign three generations to irreducible representations.
We study this point.

\begin{table}[H]
\begin{center}
\renewcommand{\arraystretch}{1.1}
\begin{tabular}{|c|c|} \hline
 group & irreducible representations \\ \hline
$A_4$ &   $1,1',1',3$ \\ \hline
$S_4$ & $1,1',2,3,3'$ \\ \hline
$A_5$ & $1,3,3',4,5$ \\ \hline
\end{tabular}
\end{center}
\caption{Irreducible representations of $A_4$, $S_4$, and $A_5$.}
\label{tab:representations}
\end{table}

\begin{table}[H]
\begin{center}
\renewcommand{\arraystretch}{1.1}
\begin{tabular}{|c|c|c|} \hline
  $k$ & $d_k$ & $A_4$ representations \\ \hline
  2 & 3 & 3 \\ \hline
  4 & 5 & $3+1+1'$ \\ \hline
  6 & 7 & $3+3+1$ \\ \hline
  8 & 9 & $3+3+1+1'+1''$ \\ \hline
  10 & 11 & $3+3+3+1+1'$ \\ \hline
\end{tabular}
\end{center}
\caption{$A_4$ representations for each weight $k$.
$d_k=k+1$.
${\bf r}=1,1',1'',3$.}
\label{tab:A4-form}
\end{table}
\begin{table}[H]
\begin{center}
\renewcommand{\arraystretch}{1.1}
\begin{tabular}{|c|c|c|} \hline
  $k$ & $d_k$ & $S_4$ representations \\ \hline
  2 & 5 & $2+3'$ \\ \hline
  4 & 9 & $1+2+3+3'$ \\ \hline
  6 & 13 & $1+1'+2+3+3'+3'$ \\ \hline
  8 & 17 & $1+2+2+3+3+3'+3'$ \\ \hline
  10 & 21 & $1+1'+2+2+3+3+3'+3'+3'$ \\ \hline
\end{tabular}
\end{center}
\caption{$S_4$ representations for each weight $k$.
$d_k=2k+1$.
${\bf r}=1,1',2,3,3'$.}
\label{tab:S4-form}
\end{table}
\begin{table}[H]
\begin{center}
\renewcommand{\arraystretch}{1.1}
\begin{tabular}{|c|c|c|} \hline
  $k$ & $d_k$ & $A_5$ representations \\ \hline
  2 & 11 & $3+3'+5$ \\ \hline
  4 & 21 & $1+3+3'+4+5+5$ \\ \hline
  6 & 31 & $1+3+3+3'+3'+4+4+5+5$ \\ \hline
\end{tabular}
\end{center}
\caption{$A_5$ representations for each weight $k$.
$d_k=5k+1$.
${\bf r}=1,3,3',4,5$.}
\label{tab:A5-form}
\end{table}

We classify the structures of the superpotential without specifying the finite modular groups.
We consider models with two pairs of Higgs fields.
They can correspond to either two singlets or a doublet.
Three generations of quarks are constructed by combining  singlets, doublets and triplets of any finite modular groups.
Table \ref{tab:AssignmentsRep} shows all possible representation combinations for up-type quarks and two pairs of up-sector Higgs fields.
\begin{table}[H]
\begin{center}
\renewcommand{\arraystretch}{1.1}
\begin{tabular}{|c|c|c|c|} \hline
  & $Q=(Q^1,Q^2,Q^3)$ & $q=(u,c,t)$ & $H_u=(H_u^1,H_u^2)$ \\ \hline
  I & singlet $\oplus$ singlet $\oplus$ singlet & singlet $\oplus$ singlet $\oplus$ singlet & \\
  II & singlet $\oplus$ singlet $\oplus$ singlet & singlet $\oplus$ doublet & \\
  II' & singlet $\oplus$ doublet & singlet $\oplus$ singlet $\oplus$ singlet & \\
  III & singlet $\oplus$ singlet $\oplus$ singlet & triplet & singlet $\oplus$ singlet \\
  III' & triplet & singlet $\oplus$ singlet $\oplus$ singlet & or \\
  IV & singlet $\oplus$ doublet & singlet $\oplus$ doublet & doublet \\
  V & singlet $\oplus$ doublet & triplet & \\
  V' & triplet & singlet $\oplus$ doublet & \\
  VI & triplet & triplet & \\ \hline
\end{tabular}
\end{center}
\caption{All possible representation combinations for up-type quarks and two pairs of  up-sector Higgs fields.}
\label{tab:AssignmentsRep}
\end{table}
In order to realize NNI forms, $Q$ and $q$ must be decomposed into $1\oplus 1_\omega \oplus 1_{\omega^2}$, and  $H_u$ must be decomposed into $1\oplus 1_\omega$, $1\oplus 1_{\omega^2}$ or $1_\omega \oplus 1_{\omega^2}$ at $\tau=\omega$, 
where ${1_{\omega^k}}$ denotes the singlet with the $Z_3^{(ST)}$ charge $k$.
Note that we need three or more independent parameters for each Higgs fields to realize the NNI form. 
We show the structures of the superpotential in each case below.
\begin{enumerate}
  \item[I.]{$Q=$ singlet $\oplus$ singlet $\oplus$ singlet, $q=$ singlet $\oplus$ singlet $\oplus$ singlet :} \\
  For $H_u={\rm singlet}\oplus{\rm singlet}$, the superpotential relevant to up-type quark mass is given by
  \begin{align}
    W &= \sum_{a=1,2}\sum_{{\bf r}_i} \left[
    {\bf Y}_{{\bf r}_i}
    \begin{pmatrix}
      Q^1 & Q^2 & Q^3 \\
    \end{pmatrix}
    \begin{pmatrix}
      \alpha_{u11}^{a{\bf r}_i} & \alpha_{u12}^{a{\bf r}_i} & \alpha_{u13}^{a{\bf r}_i} \\
      \alpha_{u21}^{a{\bf r}_i} & \alpha_{u22}^{a{\bf r}_i} & \alpha_{u23}^{a{\bf r}_i} \\
      \alpha_{u31}^{a{\bf r}_i} & \alpha_{u32}^{a{\bf r}_i} & \alpha_{u33}^{a{\bf r}_i} \\
    \end{pmatrix}
    \begin{pmatrix}
      u \\ c \\ t \\
    \end{pmatrix}
    H_u^a
    \right]_1.
  \end{align}
  For $H_u={\rm doublet}$, it is given by
  \begin{align}
  W &= \sum_{{\bf r}_i} \left[
    {\bf Y}_{{\bf r}_i}
    \begin{pmatrix}
      Q^1 & Q^2 & Q^3 \\
    \end{pmatrix}
    \begin{pmatrix}
      \alpha_{u11}^{{\bf r}_i} & \alpha_{u12}^{{\bf r}_i} & \alpha_{u13}^{{\bf r}_i} \\
      \alpha_{u21}^{{\bf r}_i} & \alpha_{u22}^{{\bf r}_i} & \alpha_{u23}^{{\bf r}_i} \\
      \alpha_{u31}^{{\bf r}_i} & \alpha_{u32}^{{\bf r}_i} & \alpha_{u33}^{{\bf r}_i} \\
    \end{pmatrix}
    \begin{pmatrix}
      u \\ c \\ t \\
    \end{pmatrix}
    H_u
    \right]_1.
  \end{align}
Both Yukawa matrices have a sufficient number of free parameters to realized the NNI form.
We show them by using $A_4$ models in the following subsection.
  \item[II.]{$Q=$ singlet $\oplus$ singlet $\oplus$ singlet, $q={\rm singlet}\oplus{\rm doublet}$:} \\
  For $H_u={\rm singlet}\oplus{\rm singlet}$, the superpotential relevant to up-type quark mass is given by
  \begin{align}
    W &= \sum_{a=1,2}\sum_{{\bf r}_i} \left[
    {\bf Y}_{{\bf r}_i}
    \begin{pmatrix}
      Q^1 & Q^2 & Q^3 \\
    \end{pmatrix}
    \begin{pmatrix}
      \alpha_{u11}^{a{\bf r}_i} & \alpha_{u12}^{a{\bf r}_i} \\
      \alpha_{u21}^{a{\bf r}_i} & \alpha_{u22}^{a{\bf r}_i} \\
      \alpha_{u31}^{a{\bf r}_i} & \alpha_{u32}^{a{\bf r}_i} \\
    \end{pmatrix}
    \begin{pmatrix}
      u \\ {\bf q}^{23} \\
    \end{pmatrix}
    H_u^a
    \right]_1, \quad
    {\bf q}^{23} = 
    \begin{pmatrix}
      c \\
      t \\
    \end{pmatrix}.
  \end{align}
  For $H_u={\rm doublet}$, it is given by
  \begin{align}
    W &= \sum_{{\bf r}_i} \left[
    {\bf Y}_{{\bf r}_i}
    \begin{pmatrix}
      Q^1 & Q^2 & Q^3 \\
    \end{pmatrix}
    \begin{pmatrix}
      \alpha_{u11}^{{\bf r}_i} & \alpha_{u12}^{{\bf r}_i} \\
      \alpha_{u21}^{{\bf r}_i} & \alpha_{u22}^{{\bf r}_i} \\
      \alpha_{u31}^{{\bf r}_i} & \alpha_{u32}^{{\bf r}_i} \\
    \end{pmatrix}
    \begin{pmatrix}
      u \\ {\bf q}^{23} \\
    \end{pmatrix}
    H_u
    \right]_1.
  \end{align}
Both Yukawa matrices have a sufficient number of free parameters to realized the NNI form.
  In the case of II', the superpotential is given by 
exchanging $Q$ and $q$ in the above.
However, these assignments can not be realized by $A_4$, $S_4$, or $A_5$ symmetry.
  \item[III.]{$Q=$ singlet $\oplus$ singlet $\oplus$ singlet, $q={\rm triplet}$:} \\
  For $H_u={\rm singlet}\oplus{\rm singlet}$, the superpotential relevant to up-type quark mass is given by
  \begin{align}
    W &= \sum_{a=1,2}\sum_{{\bf r}_i} \left[
    {\bf Y}_{{\bf r}_i}
    \begin{pmatrix}
      Q^1 & Q^2 & Q^3 \\
    \end{pmatrix}
    \begin{pmatrix}
      \alpha_{u1}^{a{\bf r}_i} \\ \alpha_{u2}^{a{\bf r}_i} \\ \alpha_{u3}^{a{\bf r}_i} \\
    \end{pmatrix}
    q
    H_u^a
    \right]_1.
  \end{align}
  For $H_u={\rm doublet}$, it is given by
  \begin{align}
    W &= \sum_{{\bf r}_i} \left[
    {\bf Y}_{{\bf r}_i}
    \begin{pmatrix}
      Q^1 & Q^2 & Q^3 \\
    \end{pmatrix}
    \begin{pmatrix}
      \alpha_{u1}^{{\bf r}_i} \\ \alpha_{u2}^{{\bf r}_i} \\ \alpha_{u3}^{{\bf r}_i} \\
    \end{pmatrix}
    q
    H_u
    \right]_1.
  \end{align}
Both Yukawa matrices have a sufficient number of free parameters to realized the NNI form.
We show them by using $A_4$ models in the following subsection.
  In the case of III', the superpotential is given by exchanging  $Q$ and $q$ in the above.
Models 1 and 2 correspond to this case.
  \item[IV.]{$Q={\rm singlet}\oplus{\rm doublet}$, $q={\rm singlet}\oplus{\rm doublet}$:} \\
  For $H_u={\rm singlet}\oplus{\rm singlet}$, the superpotential relevant to up-type quark mass is given by
  \begin{align}
    W &= \sum_{a=1,2}\sum_{{\bf r}_i} \left[
    {\bf Y}_{{\bf r}_i}
    \begin{pmatrix}
      Q^1 & {\bf Q}^{23} \\
    \end{pmatrix}
    \begin{pmatrix}
      \alpha_{u11}^{a{\bf r}_i} & \alpha_{u12}^{a{\bf r}_i} \\
      \alpha_{u21}^{a{\bf r}_i} & \alpha_{u22}^{a{\bf r}_i} \\
    \end{pmatrix}
    \begin{pmatrix}
      u \\
      {\bf q}^{23} \\
    \end{pmatrix}
    H_u^a
    \right]_1, \quad
    {\bf Q}^{23} = 
    \begin{pmatrix}
      Q^2 \\
      Q^3 \\
    \end{pmatrix}, \quad
    {\bf q}^{23} = 
    \begin{pmatrix}
      c \\
      t \\
    \end{pmatrix}.
  \end{align}
  For $H_u={\rm doublet}$, it is given by
  \begin{align}
    W &= \sum_{{\bf r}_i} \left[
    {\bf Y}_{{\bf r}_i}
    \begin{pmatrix}
      Q^1 & {\bf Q}^{23} \\
    \end{pmatrix}
    \begin{pmatrix}
      \alpha_{u11}^{{\bf r}_i} & \alpha_{u12}^{{\bf r}_i} \\
      \alpha_{u21}^{{\bf r}_i} & \alpha_{u22}^{{\bf r}_i} \\
    \end{pmatrix}
    \begin{pmatrix}
      u \\
      {\bf q}^{23} \\
    \end{pmatrix}
    H_u
    \right]_1.
  \end{align}
Both Yukawa matrices have a sufficient number of free parameters to realized the NNI form.
We show them by using $S_4$ models in the following subsection.
  \item[V.]{$Q={\rm singlet}\oplus{\rm doublet}$, $q={\rm triplet}$:} \\
  For $H_u={\rm singlet}\oplus{\rm singlet}$, the superpotential relevant to up-type quark mass is given by
  \begin{align}
    W &= \sum_{a=1,2}\sum_{{\bf r}_i} \left[
    {\bf Y}_{{\bf r}_i}
    \begin{pmatrix}
      Q^1 & {\bf Q}^{23} \\
    \end{pmatrix}
    \begin{pmatrix}
      \alpha_{u1}^{a{\bf r}_i} \\
      \alpha_{u2}^{a{\bf r}_i} \\
    \end{pmatrix}
    q
    H_u^a
    \right]_1, \quad
    {\bf Q}^{23} = 
    \begin{pmatrix}
      Q^2 \\
      Q^3 \\
    \end{pmatrix}.
  \end{align}
  For $H_u={\rm doublet}$, it is given by
  \begin{align}
    W &= \sum_{{\bf r}_i} \left[
    {\bf Y}_{{\bf r}_i}
    \begin{pmatrix}
      Q^1 & {\bf Q}^{23} \\
    \end{pmatrix}
    \begin{pmatrix}
      \alpha_{u1}^{{\bf r}_i} \\
      \alpha_{u2}^{{\bf r}_i} \\
    \end{pmatrix}
    q
    H_u
    \right]_1.
  \end{align}
  In the case of V', the superpotential is given by exchanging $Q$ and $q$ in the above.
The number of free parameters in all of these cases is insufficient to lead to the NNI form.
  \item[VI.]{$Q={\rm triplet}$, $q={\rm triplet}$:} \\
  For $H_u={\rm singlet}\oplus{\rm singlet}$, the superpotential relevant to up-type quark mass is given by
  \begin{align}
    W &= \sum_{a=1,2}\sum_{{\bf r}_i} \left[\alpha_{u}^{a{\bf r}_i}
    {\bf Y}_{{\bf r}_i}
    Q
    q
    H_u^a
    \right]_1.
  \end{align}
  For $H_u={\rm doublet}$, it is given by
  \begin{align}
    W &= \sum_{{\bf r}_i} \left[\alpha_{u}^{{\bf r}_i}
    {\bf Y}_{{\bf r}_i}
    Q
    q
    H_u
    \right]_1.
  \end{align}
\end{enumerate}
The number of free parameters in all of these cases is insufficient to lead to the NNI form.

\subsection{$A_4$ models}

The $A_4$ group has three singlets $1$, $1''$, $1'$, and a triplet $3$ as irreducible representations.
Thus, the cases I, III, III', and VI are possible.
The Yukawa couplings, which are written by modular forms, also have irreducible representations.
Table \ref{tab:A4-form} shows which representations appear  in modular forms for fixed weights.
In general, different modular forms with the same representation appears for fixed weights, 
while modular forms with some representations do not appear. 
We denote Yukawa coupling of weight $k_Y$  by
\begin{align}
  {\bf Y}_{{\bf r}_i} = ({\bf Y}_{1_j},{\bf Y}_{1''_k},{\bf Y}_{1'_\ell},{\bf Y}_{3_m}),
\end{align}
where we put indexes such as $1_j$ and $3_m$, because different modular forms with the same 
representation appear for a fixed weight.
At $\tau=\omega$, on $ST$-eigenbasis, they are transformed as
\begin{align}
  &{\bf Y}_{1_j}(\omega) \rightarrow {\bf Y}_{1_j}(ST\omega) = \omega^{2k_Y} {\bf Y}_{1_j}(\omega) , \\
  &{\bf Y}_{1''_k}(\omega) \rightarrow {\bf Y}_{1''_k}(ST\omega) = \omega^{2k_Y} \omega {\bf Y}_{1''_k}(\omega) , \\
  &{\bf Y}_{1'_\ell}(\omega) \rightarrow {\bf Y}_{1'_\ell}(ST\omega) = \omega^{2k_Y} \omega^2 {\bf Y}_{1'_\ell}(\omega) , \\
  &
  \begin{pmatrix}
    {\bf Y}_{3_m}^1(\omega) \\
    {\bf Y}_{3_m}^2(\omega) \\
    {\bf Y}_{3_m}^3(\omega) \\
  \end{pmatrix}
  \rightarrow 
  \begin{pmatrix}
    {\bf Y}_{3_m}^1(ST\omega) \\
    {\bf Y}_{3_m}^2(ST\omega) \\
    {\bf Y}_{3_m}^3(ST\omega) \\
  \end{pmatrix}
  =
  \omega^{2k_Y}
  \begin{pmatrix}
    1 & 0 & 0 \\
    0 & \omega & 0 \\
    0 & 0 & \omega^2 \\
  \end{pmatrix}
  \begin{pmatrix}
    {\bf Y}_{3_m}^1(\omega) \\
    {\bf Y}_{3_m}^2(\omega) \\
    {\bf Y}_{3_m}^3(\omega) \\
  \end{pmatrix},
\end{align}
under $ST$-transformation.
Note that $\tau = \omega$ is a fixed point of $ST$-transformation as $ST\omega=\omega$.
That means ${\bf Y}_{1_j}(ST\omega) $ must be ${\bf Y}_{1_j}(ST\omega) = {\bf Y}_{1_j}(\omega) $, and 
it vanishes unless $k_Y=6n$.
The other Yukawa couplings have similar behaviors.
Then, it is found that when $k_Y=6n$, ${\bf Y}_{1_j}(\omega)$ and ${\bf Y}_{3_m}^1(\omega)$ are non-vanishing, while 
the other Yukawa couplings vanish.
Similarly, only ${\bf Y}_{1''_k}(\omega)$ and $ {\bf Y}_{3_m}^2(\omega)$ are non-vanishing for $k_Y=6n+4$, 
while only ${\bf Y}_{1'_k}(\omega)$ and $ {\bf Y}_{3_m}^3(\omega)$ are non-vanishing for $k_Y=6n+2$. 
In what follows, we focus on the case only $({\bf Y}_{1_j}(\omega),{\bf Y}_{3_m}^1(\omega))$ remain non-vanishing, that is,
\begin{align}
  &{\bf Y}_{{\bf r}_i}(\omega) = ({\bf Y}_{1_j}(\omega),0,0,{\bf Y}_{3_m}(\omega)),
\end{align}
with
\begin{align}
  {\bf Y}_{3_m}(\omega) =
  \begin{pmatrix}
    {\bf Y}_{3_m}^1(\omega) \\
    0 \\
    0 \\
  \end{pmatrix},
\end{align}
although we can discuss other choices similarly.

Since the $ST$-invariances restrict Yukawa couplings, the structures of the mass matrix as well as the superpotential are also constrained.
For example, let us study the superpotential and the mass matrix in the case I.
In this case, quark doublet and right-handed up-type quark singlets are assigned into $(1,1'',1')$; up-sector  Higgs fields are assigned into either of $(1,1'')$, $(1,1')$ or $(1'',1')$.
We focus on up-sector Higgs fields with $(1,1'')$.
Then the superpotential at $\tau=\omega$ is given by
\begin{align}
  W (\omega)&= \left[
    \begin{pmatrix}
      Q^1 & Q^2 & Q^3 \\
    \end{pmatrix}
    \sum_j {\bf Y}_{1_j}(\omega)
    \begin{pmatrix}
      \alpha_{u11}^{11_j} & \alpha_{u12}^{11_j} & \alpha_{u13}^{11_j} \\
      \alpha_{u21}^{11_j} & \alpha_{u22}^{11_j} & \alpha_{u23}^{11_j} \\
      \alpha_{u31}^{11_j} & \alpha_{u32}^{11_j} & \alpha_{u33}^{11_j} \\
    \end{pmatrix}
    \begin{pmatrix}
      u \\ c \\ t \\
    \end{pmatrix}
    \right]_1 H_u^1 \notag \\
    &+ \left[
    \begin{pmatrix}
      Q^1 & Q^2 & Q^3 \\
    \end{pmatrix}
    \sum_j {\bf Y}_{1_j}(\omega)
    \begin{pmatrix}
      \alpha_{u11}^{21_j} & \alpha_{u12}^{21_j} & \alpha_{u13}^{21_j} \\
      \alpha_{u21}^{21_j} & \alpha_{u22}^{21_j} & \alpha_{u23}^{21_j} \\
      \alpha_{u31}^{21_j} & \alpha_{u32}^{21_j} & \alpha_{u33}^{21_j} \\
    \end{pmatrix}
    \begin{pmatrix}
      u \\ c \\ t \\
    \end{pmatrix}
    \right]_{1'} H_u^2 \notag \\
    &+ \left[
    \begin{pmatrix}
      Q^1 & Q^2 & Q^3 \\
    \end{pmatrix}
    \sum_m {\bf Y}_{3_m}(\omega)
    \begin{pmatrix}
      \alpha_{u11}^{13_m} & \alpha_{u12}^{13_m} & \alpha_{u13}^{13_m} \\
      \alpha_{u21}^{13_m} & \alpha_{u22}^{13_m} & \alpha_{u23}^{13_m} \\
      \alpha_{u31}^{13_m} & \alpha_{u32}^{13_m} & \alpha_{u33}^{13_m} \\
    \end{pmatrix}
    \begin{pmatrix}
      u \\ c \\ t \\
    \end{pmatrix}
    \right]_1 H_u^1 \notag \\
    &+ \left[
    \begin{pmatrix}
      Q^1 & Q^2 & Q^3 \\
    \end{pmatrix}
    \sum_m {\bf Y}_{3_m}(\omega)
    \begin{pmatrix}
      \alpha_{u11}^{23_m} & \alpha_{u12}^{23_m} & \alpha_{u13}^{23_m} \\
      \alpha_{u21}^{23_m} & \alpha_{u22}^{23_m} & \alpha_{u23}^{23_m} \\
      \alpha_{u31}^{23_m} & \alpha_{u32}^{23_m} & \alpha_{u33}^{23_m} \\
    \end{pmatrix}
    \begin{pmatrix}
      u \\ c \\ t \\
    \end{pmatrix}
    \right]_{1'} H_u^2.
\end{align}
By use of the multiplication rule in Appendix \ref{app:A4group}, it follows that
\begin{align}
  W(\omega) &= 
  \begin{pmatrix}
      Q^1 & Q^2 & Q^3 \\
    \end{pmatrix}
    \sum_j {\bf Y}_{1_j}(\omega)
    \begin{pmatrix}
      \alpha_{u11}^{11_j} & 0 & 0 \\
      0 & 0 & \alpha_{u23}^{11_j} \\
      0 & \alpha_{u32}^{11_j} & 0 \\
    \end{pmatrix}
    \begin{pmatrix}
      u \\ c \\ t \\
    \end{pmatrix} H_u^1 \notag \\
    &+
    \begin{pmatrix}
      Q^1 & Q^2 & Q^3 \\
    \end{pmatrix}
    \sum_j {\bf Y}_{1_j}(\omega)
    \begin{pmatrix}
      0 & 0 & \alpha_{u13}^{21_j} \\
      0 & \alpha_{u22}^{21_j} & 0 \\
      \alpha_{u31}^{21_j} & 0 & 0 \\
    \end{pmatrix}
    \begin{pmatrix}
      u \\ c \\ t \\
    \end{pmatrix} H_u^2.
\end{align}
The up-type quark mass matrix is obtained as
\begin{align}
  M_u &=
  \sum_j {\bf Y}_{1_j}(\omega)
  \begin{pmatrix}
    \alpha_{u11}^{11_j} & 0 & 0 \\
    0 & 0 & \alpha_{u23}^{11_j} \\
    0 & \alpha_{u32}^{11_j} & 0 \\
  \end{pmatrix}
  \langle H_u^1 \rangle
  +
  \sum_j {\bf Y}_{1_j}(\omega)
  \begin{pmatrix}
    0 & 0 & \alpha_{u13}^{21_j} \\
    0 & \alpha_{u22}^{21_j} & 0 \\
    \alpha_{u31}^{21_j} & 0 & 0 \\
  \end{pmatrix}
  \langle H_u^2 \rangle.
\end{align}
Note that all nonzero elements in the mass matrix are written by independent parameters.
Therefore we can realize NNI forms in the case I.

Next, let us study the mass matrix and the superpotential in the case III.
In this case, quark doublet is assigned into $(1,1'',1)$; right-handed up-type quark singlets are assigned into the triplet $3$; up-sector Higgs fields are assigned into either of $(1,1'')$, $(1,1')$ or $(1'',1')$.
We focus on up-sector Higgs fields with $(1,1'')$.
Then the superpotential at $\tau=\omega$ is given by
\begin{align}
    W (\omega)&= \left[
    \sum_j {\bf Y}_{1_j}
    \begin{pmatrix}
      Q^1 & Q^2 & Q^3 \\
    \end{pmatrix}
    \begin{pmatrix}
      \alpha_{u1}^{11_j} \\ \alpha_{u2}^{11_j} \\ \alpha_{u3}^{11_j} \\
    \end{pmatrix}
    q
    \right]_1 H_u^1
    + \left[
    \sum_j {\bf Y}_{1_j}
    \begin{pmatrix}
      Q^1 & Q^2 & Q^3 \\
    \end{pmatrix}
    \begin{pmatrix}
      \alpha_{u1}^{21_j} \\ \alpha_{u2}^{21_j} \\ \alpha_{u3}^{21_j} \\
    \end{pmatrix}
    q
    \right]_{1'} H_u^2 \notag \\
    &+ \left[
    \sum_m {\bf Y}_{3_m}
    \begin{pmatrix}
      Q^1 & Q^2 & Q^3 \\
    \end{pmatrix}
    \begin{pmatrix}
      \alpha_{u1}^{13_m} \\ \alpha_{u2}^{13_m} \\ \alpha_{u3}^{13_m} \\
    \end{pmatrix}
    q
    \right]_1 H_u^1
    + \left[
    \sum_m {\bf Y}_{3_m}
    \begin{pmatrix}
      Q^1 & Q^2 & Q^3 \\
    \end{pmatrix}
    \begin{pmatrix}
      \alpha_{u1}^{23_m} \\ \alpha_{u2}^{23_m} \\ \alpha_{u3}^{23_m} \\
    \end{pmatrix}
    q
    \right]_{1'} H_u^2.
\end{align}
By use of the multiplication rule in Appendix \ref{app:A4group}, it follows that
\begin{align}
    W(\omega) &=
    \begin{pmatrix}
      Q^1 & Q^2 & Q^3 \\
    \end{pmatrix}
    \omega^2\sum_m {\bf Y}_{3_m}^1
    \begin{pmatrix}
      \alpha_{u1}^{13_m} & 0 & 0 \\
      0 & 0 & \alpha_{u2}^{13_m} \\
      0 & \alpha_{u3}^{13_m} & 0 \\
    \end{pmatrix}
    \begin{pmatrix}
      u \\ c \\ t \\
    \end{pmatrix}
    H_u^1 \notag \\
    &+
    \begin{pmatrix}
      Q^1 & Q^2 & Q^3 \\
    \end{pmatrix}
    \omega^2\sum_m {\bf Y}_{3_m}^1
    \begin{pmatrix}
      0 & 0 & \alpha_{u1}^{23_m} \\
      0 & \alpha_{u2}^{23_m} & 0 \\
      \alpha_{u3}^{23_m} & 0 & 0 \\
    \end{pmatrix}
    \begin{pmatrix}
      u \\ c \\ t \\
    \end{pmatrix}
    H_u^2.
\end{align}
The up-type quark mass matrix is obtained as
\begin{align}
    M_u &=
    \omega^2\sum_m {\bf Y}_{3_m}^1
    \begin{pmatrix}
      \alpha_{u1}^{13_m} & 0 & 0 \\
      0 & 0 & \alpha_{u2}^{13_m} \\
      0 & \alpha_{u3}^{13_m} & 0 \\
    \end{pmatrix} \langle H_u^1 \rangle
    +
    \omega^2\sum_m {\bf Y}_{3_m}^1
    \begin{pmatrix}
      0 & 0 & \alpha_{u1}^{23_m} \\
      0 & \alpha_{u2}^{23_m} & 0 \\
      \alpha_{u3}^{23_m} & 0 & 0 \\
    \end{pmatrix} \langle H_u^2 \rangle .
  \end{align}
All nonzero elements in this mass matrix are obtained by independent parameters.
Therefore we can realize NNI forms in the case III.
Also, it can be realized in the case III'.
Indeed, models 1 and 2 correspond to this case.

As a result, NNI forms can be found in only the cases I, III and III'.
However, the number of free parameters in the case VI is not large enough to 
realize the NNI form.


\subsection{$S_4$ models}

The $S_4$ group has two singlets, 1, $1'$, a doublet $2$, and two triplets $3$ and $3'$.
Thus, cases IV, V, V', and VI are possible.
In $S_4$ model, Yukawa couplings of weight $k_Y$ are denoted by
\begin{align}
  {\bf Y}_{{\bf r}_i} = ({\bf Y}_{1_j},{\bf Y}_{1'_k},{\bf Y}_{2_\ell},{\bf Y}_{3_m},{\bf Y}_{3'_n}).
\end{align}
At $\tau=\omega$, on $ST$-eigenbasis, they are transformed as
\begin{align}
  &{\bf Y}_{1_j}(\omega) \rightarrow {\bf Y}_{1_j}(ST\omega) = \omega^{2k_Y} {\bf Y}_{1_j}(\omega) , \\
  &{\bf Y}_{1'_k}(\omega) \rightarrow {\bf Y}_{1'_k}(ST\omega) = \omega^{2k_Y} {\bf Y}_{1'_k}(\omega) , \\
  &
  \begin{pmatrix}
    {\bf Y}_{2_\ell}^1(\omega) \\
    {\bf Y}_{2_\ell}^2(\omega) \\
  \end{pmatrix}
  \rightarrow
  \begin{pmatrix}
    {\bf Y}_{2_\ell}^1(ST\omega) \\
    {\bf Y}_{2_\ell}^2(ST\omega) \\
  \end{pmatrix}
  = \omega^{2k_Y}
  \begin{pmatrix}
    \omega & 0 \\
    0 & \omega^2 \\
  \end{pmatrix}
  \begin{pmatrix}
    {\bf Y}_{2_\ell}^1(\omega) \\
    {\bf Y}_{2_\ell}^2(\omega) \\
  \end{pmatrix}
\\
  &
  \begin{pmatrix}
    {\bf Y}_{3_m}^1(\omega) \\
    {\bf Y}_{3_m}^2(\omega) \\
    {\bf Y}_{3_m}^3(\omega) \\
  \end{pmatrix}
  \rightarrow 
  \begin{pmatrix}
    {\bf Y}_{3_m}^1(ST\omega) \\
    {\bf Y}_{3_m}^2(ST\omega) \\
    {\bf Y}_{3_m}^3(ST\omega) \\
  \end{pmatrix}
  =
  \omega^{2k_Y}
  \begin{pmatrix}
    1 & 0 & 0 \\
    0 & \omega & 0 \\
    0 & 0 & \omega^2 \\
  \end{pmatrix}
  \begin{pmatrix}
    {\bf Y}_{3_m}^1(\omega) \\
    {\bf Y}_{3_m}^2(\omega) \\
    {\bf Y}_{3_m}^3(\omega) \\
  \end{pmatrix}, \\
  &
  \begin{pmatrix}
    {\bf Y}_{3'_n}^1(\omega) \\
    {\bf Y}_{3'_n}^2(\omega) \\
    {\bf Y}_{3'_n}^3(\omega) \\
  \end{pmatrix}
  \rightarrow 
  \begin{pmatrix}
    {\bf Y}_{3'_n}^1(ST\omega) \\
    {\bf Y}_{3'_n}^2(ST\omega) \\
    {\bf Y}_{3'_n}^3(ST\omega) \\
  \end{pmatrix}
  =
  \omega^{2k_Y}
  \begin{pmatrix}
    1 & 0 & 0 \\
    0 & \omega & 0 \\
    0 & 0 & \omega^2 \\
  \end{pmatrix}
  \begin{pmatrix}
    {\bf Y}_{3'_n}^1(\omega) \\
    {\bf Y}_{3'_n}^2(\omega) \\
    {\bf Y}_{3'_n}^3(\omega) \\
  \end{pmatrix},
\end{align}
under $ST$-transformation.
As in $A_4$ models, $ST$-invariances of Yukawa couplings at $\tau=\omega$ mean 
that for $k_Y=6n$, the following Yukawa couplings 
\begin{align}
  \left({\bf Y}_{1_j}(\omega),{\bf Y}_{1'_k}(\omega),{\bf Y}_{3_m}^1(\omega),{\bf Y}_{3'_n}^1(\omega)\right),
\end{align}
remain non-vanishing at $\tau = \omega$, while the others vanish.
For $k_Y=6n+4$, only the following Yukawa couplings
\begin{align}
\left({\bf Y}_{2_\ell}^1(\omega),{\bf Y}_{3_m}^2(\omega),{\bf Y}_{3'_n}^2(\omega)\right),
\end{align}
remain non-vanishing, while for $k_Y=6n+2$, 
only the following Yukawa couplings 
\begin{align}
\left({\bf Y}_{2_\ell}^2(\omega),{\bf Y}_{3_m}^3(\omega),{\bf Y}_{3'_n}^3(\omega)\right),
\end{align}
remain non-vanishing.
In what follows, we focus on the case only $({\bf Y}_{1_j}(\omega),{\bf Y}_{1'_k}(\omega),{\bf Y}_{3_m}^1(\omega),{\bf Y}_{3'_n}^1(\omega))$ 
remain, that is,
\begin{align}
  &{\bf Y}_{{\bf r}_i}(\omega) = ({\bf Y}_{1_j}(\omega),{\bf Y}_{1'_k}(\omega),0,{\bf Y}_{3_m}(\omega),{\bf Y}_{3'_n}(\omega)),
\end{align}
with
\begin{align}
  {\bf Y}_{3_m}(\omega) =
  \begin{pmatrix}
    {\bf Y}_{3_m}^1(\omega) \\
    0 \\
    0 \\
  \end{pmatrix}, \quad
  {\bf Y}_{3'_n}(\omega) =
  \begin{pmatrix}
    {\bf Y}_{3'_n}^1(\omega) \\
    0 \\
    0 \\
  \end{pmatrix}.
\end{align}
Note that in this choice we have four kinds of nonzero Yukawa couplings, that is, four independent parameters corresponding to irreducible representations of $S_4$, $1,1',3$ and $3'$.
In the other choices, we never realize NNI forms in $S_4$ models because  the number of free parameters is not sufficiently large.

To find NNI forms, we consider the superpotential in the case IV.
In this case, quark doublet and right-handed up-type quark singlets are assigned into $(1,2)$ or $(1',2)$; up-sector Higgs fields are assigned into the doublet $2$.
We focus on quark doublet with $(1,2)$ and right-handed up-type quark singlets with $(1,2)$.
Then the superpotential at $\tau=\omega$ is given by
\begin{align}
  W(\omega) &=
  \left[
  \begin{pmatrix}
    Q^1 & {\bf Q}^{23} \\
  \end{pmatrix}
  \sum_j {\bf Y}_{1_j}(\omega)
  \begin{pmatrix}
    \alpha_{u11}^{1_j} & \alpha_{u12}^{1_j} \\
    \alpha_{u21}^{1_j} & \alpha_{u22}^{1_j} \\
  \end{pmatrix}
  \begin{pmatrix}
    u \\ {\bf q}^{23} \\
  \end{pmatrix}
  H_u
  \right]_1 \notag \\
  &+ \left[
  \begin{pmatrix}
    Q^1 & {\bf Q}^{23} \\
  \end{pmatrix}
  \sum_k {\bf Y}_{1'_k}(\omega)
  \begin{pmatrix}
    \alpha_{u11}^{1'_k} & \alpha_{u12}^{1'_k} \\
    \alpha_{u21}^{1'_k} & \alpha_{u22}^{1'_k} \\
  \end{pmatrix}
  \begin{pmatrix}
    u \\ {\bf q}^{23} \\
  \end{pmatrix}
  H_u
  \right]_1 \notag \\
  &+ \left[
  \begin{pmatrix}
    Q^1 & {\bf Q}^{23} \\
  \end{pmatrix}
  \sum_m {\bf Y}_{3_m}(\omega)
  \begin{pmatrix}
    \alpha_{u11}^{3_m} & \alpha_{u12}^{3_m} \\
    \alpha_{u21}^{3_m} & \alpha_{u22}^{3_m} \\
  \end{pmatrix}
  \begin{pmatrix}
    u \\ {\bf q}^{23} \\
  \end{pmatrix}
  H_u
  \right]_1 \notag \\
  &+ \left[
  \begin{pmatrix}
    Q^1 & {\bf Q}^{23} \\
  \end{pmatrix}
  \sum_n {\bf Y}_{3'_n}(\omega)
  \begin{pmatrix}
    \alpha_{u11}^{3'_n} & \alpha_{u12}^{3'_n} \\
    \alpha_{u21}^{3'_n} & \alpha_{u22}^{3'_n} \\
  \end{pmatrix}
  \begin{pmatrix}
    u \\ {\bf q}^{23} \\
  \end{pmatrix}
  H_u
  \right]_1.
\end{align}
By use of the multiplication rule in  Appendix \ref{app:S4group}, it follows that
\begin{align}
  W(\omega) =
  \begin{pmatrix}
    Q^1 & Q^2 & Q^3 \\
  \end{pmatrix}
  &\left[
  -\sum_j\frac{{\bf Y}_{1_j}}{\sqrt{2}} \alpha_{u21}^{1_j}
  \begin{pmatrix}
    0 & 0 & 0 \\
    H_u^2 & 0 & 0 \\
    H_u^1 & 0 & 0 \\
  \end{pmatrix} 
  -\sum_j\frac{{\bf Y}_{1_j}}{\sqrt{2}} \alpha_{u12}^{1_j}
  \begin{pmatrix}
    0 & H_u^2 & H_u^1 \\
    0 & 0 & 0 \\
    0 & 0 & 0 \\
  \end{pmatrix} \right. \notag \\
  &+\sum_k\frac{i{\bf Y}_{1'_k}}{\sqrt{2}} \alpha_{u21}^{1'_k}
  \begin{pmatrix}
    0 & 0 & 0 \\
    -H_u^2 & 0 & 0 \\
    H_u^1 & 0 & 0 \\
  \end{pmatrix}
  +\sum_k\frac{i{\bf Y}_{1'_k}}{\sqrt{2}} \alpha_{u12}^{1'_k}
  \begin{pmatrix}
    0 & -H_u^2 & H_u^1 \\
    0 & 0 & 0 \\
    0 & 0 & 0 \\
  \end{pmatrix} \notag \\
  &\left.
  +\sum_j\frac{i{\bf Y}_{1_j}}{2} \alpha_{u22}^{1_j}
  \begin{pmatrix}
    0 & 0 & 0 \\
    0 & H_u^1 & 0 \\
    0 & 0 & H_u^2 \\
  \end{pmatrix}
  +\sum_k\frac{{\bf Y}_{1'_k}}{2} \alpha_{u22}^{1'_k}
  \begin{pmatrix}
    0 & 0 & 0 \\
    0 & H_u^1 & 0 \\
    0 & 0 & -H_u^2 \\
  \end{pmatrix}
  \right]
  \begin{pmatrix}
    u \\ c \\ t \\
  \end{pmatrix}.
\end{align}
Defining
\begin{align}
  &\alpha^1 \equiv -\sum_j \frac{{\bf Y}_{1_j}}{\sqrt{2}} \alpha_{u21}^{1_j}+\sum_k \frac{i{\bf Y}_{1'_k}}{\sqrt{2}} \alpha_{u21}^{1'_k}, 
  \quad \alpha^2 \equiv -\sum_j \frac{{\bf Y}_{1_j}}{\sqrt{2}} \alpha_{u21}^{1_j}-\sum_k \frac{i{\bf Y}_{1'_k}}{\sqrt{2}} \alpha_{u21}^{1'_k}, \\
  &\beta^1 \equiv -\sum_j \frac{{\bf Y}_{1_j}}{\sqrt{2}} \alpha_{u12}^{1_j}+\sum_k \frac{i{\bf Y}_{1'_k}}{\sqrt{2}} \alpha_{u12}^{1'_k}, 
  \quad \beta^2 \equiv -\sum_j \frac{{\bf Y}_{1_j}}{\sqrt{2}} \alpha_{u12}^{1_j}-\sum_k \frac{i{\bf Y}_{1'_k}}{\sqrt{2}} \alpha_{u12}^{1'_k}, \\
  &\gamma^1 \equiv \sum_j \frac{i{\bf Y}_{1_j}}{2} \alpha_{u22}^{1_j}+\sum_k \frac{{\bf Y}_{1'_k}}{2} \alpha_{u22}^{1'_k}, 
  \quad \gamma^2 \equiv \sum_j \frac{i{\bf Y}_{1_j}}{2} \alpha_{u22}^{1_j}-\sum_k \frac{{\bf Y}_{1'_k}}{2} \alpha_{u22}^{1'_k},
\end{align}
the up-type quark mass matrix is obtained as
\begin{align}
  M_u &=
  \begin{pmatrix}
    0 & 0 & \beta^1 \\
    0 & \gamma^1 & 0 \\
    \alpha^1 & 0 & 0 \\
  \end{pmatrix} \langle H_u^1 \rangle
  +
  \begin{pmatrix}
    0 & \beta^2 & 0 \\
    \alpha^2 & 0 & 0 \\
    0 & 0 & \gamma^2 \\
  \end{pmatrix} \langle H_u^2 \rangle.
\end{align}
All nonzero elements in this mass matrix can be written by the parameters $\alpha^{1,2},\beta^{1,2}$ and $\gamma^{1,2}$, 
which are independent.
Therefore we can realize NNI forms in the case IV.

In a similar way, we can find the structures of the mass matrices in all cases.
The number of free parameters is not sufficient in cases V, V', and VI in order to lead to the NNI form.
As a result, NNI forms can be found in only the case IV.


\subsection{$A_5$ models}

In $A_5$ group, there is only one singlet $1$ and no doublets.
Thus two pairs of Higgs fields cannot have different $ST$ charges and we never find NNI forms.
Table \ref{tab:NNIinA4S4A5} shows our results in models with two pairs of Higgs fields.
\begin{table}[H]
\begin{center}
\renewcommand{\arraystretch}{1.1}
\begin{tabular}{|c|c|c|c|} \hline
  & $A_4$ & $S_4$ & $A_5$ \\ \hline
  NNI forms & I, III, III' & IV & None \\ \hline
\end{tabular}
\end{center}
\caption{NNI forms in $A_4$, $S_4$  and $A_5$ models with two pairs of Higgs fields.}
\label{tab:NNIinA4S4A5}
\end{table}

On the other hand, we can realize the NNI form in $A_5$ modular symmetric models with four pairs of Higgs fields\footnote{In the models with three pairs of Higgs fields, we can find Higgs fields with three different $ST$-charges but the number of free parameters in such models is not large enough to realize NNI forms.}.
The $A_5$ group has one singlet $1$, two triplets $3$ and $3'$, one four-dimensional representation $4$ and one five-dimensional representation $5$.
In $A_5$ model, Yukawa couplings of weight $k_Y$ are denoted by
\begin{align}
  {\bf Y}_{{\bf r}_i} = ({\bf Y}_{1_j}, {\bf Y}_{3_k}, {\bf Y}_{3'_\ell}, {\bf Y}_{4_m}, {\bf Y}_{5_n}).
\end{align}
At $\tau=\omega$, on $ST$-eigenbasis, they are transformed as
\begin{align}
  &{\bf Y}_{1_j}(\omega) \rightarrow {\bf Y}_{1_j}(ST\omega) = \omega^{2k_Y} {\bf Y}_{1_j}(\omega), \\
  &\begin{pmatrix} {\bf Y}_{3_k}^1(\omega)\\ {\bf Y}_{3_k}^2(\omega)\\ {\bf Y}_{3_k}^3(\omega)\\ \end{pmatrix} \rightarrow \begin{pmatrix} {\bf Y}_{3_k}^1(ST\omega)\\ {\bf Y}_{3_k}^2(ST\omega)\\ {\bf Y}_{3_k}^3(ST\omega)\\ \end{pmatrix} = \omega^{2k_Y} \begin{pmatrix}1&0&0\\0&\omega&0\\0&0&\omega^2\\\end{pmatrix} \begin{pmatrix} {\bf Y}_{3_k}^1(\omega)\\ {\bf Y}_{3_k}^2(\omega)\\ {\bf Y}_{3_k}^3(\omega)\\ \end{pmatrix}, \\
  &\begin{pmatrix} {\bf Y}_{3'_\ell}^1(\omega)\\ {\bf Y}_{3'_\ell}^2(\omega)\\ {\bf Y}_{3'_\ell}^3(\omega)\\ \end{pmatrix} \rightarrow \begin{pmatrix} {\bf Y}_{3'_\ell}^1(ST\omega)\\ {\bf Y}_{3'_\ell}^2(ST\omega)\\ {\bf Y}_{3'_\ell}^3(ST\omega)\\ \end{pmatrix} = \omega^{2k_Y} \begin{pmatrix}1&0&0\\0&\omega&0\\0&0&\omega^2\\\end{pmatrix} \begin{pmatrix} {\bf Y}_{3'_\ell}^1(\omega)\\ {\bf Y}_{3'_\ell}^2(\omega)\\ {\bf Y}_{3'_\ell}^3(\omega)\\ \end{pmatrix}, \\
  &\begin{pmatrix} {\bf Y}_{4_m}^1(\omega)\\ {\bf Y}_{4_m}^2(\omega)\\ {\bf Y}_{4_m}^3(\omega)\\ {\bf Y}_{4_m}^4(\omega)\\ \end{pmatrix} \rightarrow \begin{pmatrix} {\bf Y}_{4_m}^1(ST\omega)\\ {\bf Y}_{4_m}^2(ST\omega)\\ {\bf Y}_{4_m}^3(ST\omega)\\ {\bf Y}_{4_m}^4(ST\omega)\\ \end{pmatrix} = \omega^{2k_Y} \begin{pmatrix}1&0&0&0\\0&1&0&0\\0&0&\omega&0\\0&0&0&\omega^2\end{pmatrix} \begin{pmatrix} {\bf Y}_{4_m}^1(\omega)\\ {\bf Y}_{4_m}^2(\omega)\\ {\bf Y}_{4_m}^3(\omega)\\ {\bf Y}_{4_m}^4(\omega)\\ \end{pmatrix}, \\
  &\begin{pmatrix} {\bf Y}_{5_n}^1(\omega)\\ {\bf Y}_{5_n}^2(\omega)\\ {\bf Y}_{5_n}^3(\omega)\\ {\bf Y}_{5_n}^4(\omega)\\ {\bf Y}_{5_n}^5(\omega)\\ \end{pmatrix} \rightarrow \begin{pmatrix} {\bf Y}_{5_n}^1(ST\omega)\\ {\bf Y}_{5_n}^2(ST\omega)\\ {\bf Y}_{5_n}^3(ST\omega)\\ {\bf Y}_{5_n}^4(ST\omega)\\ {\bf Y}_{5_n}^5(ST\omega)\\ \end{pmatrix} = \omega^{2k_Y} \begin{pmatrix}1&0&0&0&0\\0&\omega&0&0&0\\0&0&\omega&0&0\\0&0&0&\omega^2&0\\0&0&0&0&\omega^2\\\end{pmatrix} \begin{pmatrix} {\bf Y}_{5_n}^1(\omega)\\ {\bf Y}_{5_n}^2(\omega)\\ {\bf Y}_{5_n}^3(\omega)\\ {\bf Y}_{5_n}^4(\omega)\\ {\bf Y}_{5_n}^5(\omega)\\ \end{pmatrix},
\end{align}
under $ST$-transformation.
As in $A_4$ and $S_4$ models, $ST$-invariances of Yukawa couplings at $\tau=\omega$ mean that for $k_Y=6n$, the following Yukawa couplings
\begin{align}
  ({\bf Y}_{1_j}(\omega), {\bf Y}_{3_k}^1(\omega), {\bf Y}_{3'_\ell}^1(\omega), {\bf Y}_{4_m}^1(\omega), {\bf Y}_{4_m}^2(\omega), {\bf Y}_{5_n}^1(\omega)),
\end{align}
remain non-vanishing at $\tau=\omega$, while the other vanish.
For $k_Y=6n+4$, only the following Yukawa couplings
\begin{align}
  ({\bf Y}_{3_k}^2(\omega), {\bf Y}_{3'_\ell}^2(\omega), {\bf Y}_{4_m}^3(\omega), {\bf Y}_{5_n}^2(\omega), {\bf Y}_{5_n}^3(\omega)),
\end{align}
remain non-vanishing at $\tau=\omega$, while for $k_Y=6n+2$, only the following Yukawa couplings
\begin{align}
  ({\bf Y}_{3_k}^3(\omega), {\bf Y}_{3'_\ell}^3(\omega), {\bf Y}_{4_m}^4(\omega), {\bf Y}_{5_n}^4(\omega), {\bf Y}_{5_n}^5(\omega)),
\end{align}
remain non-vanishing.
In what follows, we focus on the case only (${\bf Y}_{3_k}^2(\omega)$, ${\bf Y}_{3'_\ell}^2(\omega)$, ${\bf Y}_{4_m}^3(\omega)$, ${\bf Y}_{5_n}^2(\omega)$, ${\bf Y}_{5_n}^3(\omega)$) remain, that is,
\begin{align}
  {\bf Y}_{{\bf r}_i}(\omega) = (0, {\bf Y}_{3_k}(\omega), {\bf Y}_{3'_\ell}(\omega), {\bf Y}_{4_m}(\omega), {\bf Y}_{5_n}(\omega)) ,
\end{align}
with
\begin{align}
  &{\bf Y}_{3_k}(\omega) = \begin{pmatrix} 0\\ {\bf Y}_{3_k}^2(\omega) \\ 0\\ \end{pmatrix}, \quad {\bf Y}_{3'_\ell}(\omega) = \begin{pmatrix} 0\\ {\bf Y}_{3'_\ell}^2(\omega) \\ 0\\ \end{pmatrix}, \\
  &{\bf Y}_{4_m}(\omega) = \begin{pmatrix} 0\\ 0\\ {\bf Y}_{4_m}^3(\omega) \\ 0 \\ \end{pmatrix}, \quad {\bf Y}_{5_n}(\omega) = \begin{pmatrix} 0\\ {\bf Y}_{5_n}^2(\omega) \\ {\bf Y}_{5_n}^3(\omega) \\ 0 \\ 0 \\ \end{pmatrix},
\end{align}
although we can discuss other choices similarly.
Note that in this choice we have four kinds of nonzero Yukawa couplings, that is, four independent parameters corresponding to irreducible representations of $A_5$, $3$, $3'$, $4$ and $5$.

Next we consider the superpotential.
Since Higgs fields must contain two different $ST$-charge modes to realize the NNI forms, four pairs of Higgs fields should be assigned into one singlet and one triplet, $(1,3)$ or $(1,3')$.
Similarly, quark doublet and right-handed up-type quark singlets must contain three different $ST$-charge modes and they should be assigned into triplets, $3$ or $3'$.
We focus on quark doublet with $3$, right-handed up-type quark singlets with $3$ and up-sector Higgs fields with $(1,3)$.
Then the superpotential at $\tau=\omega$ is given by
\begin{align}
  W(\omega) &= \left[Qq \sum_j{\bf Y}_{1_j}(\omega)\begin{pmatrix} \alpha_{u1}^{1_j} & \alpha_{u2}^{1_j} \\ \end{pmatrix} \begin{pmatrix} H_u^1 \\ {\bf H_u} \\ \end{pmatrix} \right]_1
  +  \left[Qq \sum_k{\bf Y}_{3_k}(\omega)\begin{pmatrix} \alpha_{u1}^{3_k} & \alpha_{u2}^{3_k} \\ \end{pmatrix} \begin{pmatrix} H_u^1 \\ {\bf H_u} \\ \end{pmatrix} \right]_1 \notag \\
  &+ \left[Qq\sum_\ell{\bf Y}_{3'_\ell}(\omega)\begin{pmatrix} \alpha_{u1}^{3'_\ell} & \alpha_{u2}^{3'_\ell} \\ \end{pmatrix} \begin{pmatrix} H_u^1 \\ {\bf H_u} \\ \end{pmatrix} \right]_1
  + \left[Qq\sum_m{\bf Y}_{4_m}(\omega)\begin{pmatrix} \alpha_{u1}^{4_m} & \alpha_{u2}^{4_m} \\ \end{pmatrix} \begin{pmatrix} H_u^1 \\ {\bf H_u} \\ \end{pmatrix} \right]_1 \notag \\
  &+ \left[Qq\sum_n{\bf Y}_{5_n}(\omega)\begin{pmatrix} \alpha_{u1}^{5_n} & \alpha_{u2}^{5_n} \\ \end{pmatrix} \begin{pmatrix} H_u^1 \\ {\bf H_u} \\ \end{pmatrix} \right]_1,
\end{align}
where
\begin{align}
  {\bf H_u} &\equiv
  \begin{pmatrix}
    H_u^2 \\ H_u^3 \\ H_u^4 \\
  \end{pmatrix}.
\end{align}
By use of the multiplication rule in Appendix \ref{app:A5group}, it follows that
\begin{align}
  W(\omega) &= 
  \begin{pmatrix}
    Q^1 & Q^2 & Q^3 \\
  \end{pmatrix}
  \left[
  \sum_k \alpha_{u1}^{3_k}{\bf Y}_{3_k}^2(\omega) 
  \begin{pmatrix}
    0 & 0 & -1 \\
    0 & 0 & 0 \\
    1 & 0 & 0 \\
  \end{pmatrix}
  H_u^1 
  \right.\notag \\
  &+
  \sum_k \alpha_{u2}^{3_k}{\bf Y}_{3_k}^2(\omega) \left(
  \begin{pmatrix}
    0 & 0 & -4 \\
    0 & 0 & 0 \\
    -2 & 0 & 0 \\
  \end{pmatrix}
  H_u^2 +
  \begin{pmatrix}
    0 & 0 & 0 \\
    0 & 0 & 0 \\
    0 & 0 & -6 \\
  \end{pmatrix}
  H_u^3 +
  \begin{pmatrix}
    3 & 0 & 0 \\
    0 & 0 & -1 \\
    0 & 1 & 0 \\
  \end{pmatrix}
  H_u^4 \right) \notag \\
  &+\sum_\ell \frac{\alpha_{u2}^{3'_\ell}}{\sqrt{3}}{\bf Y}_{3'_\ell}^2(\omega)
  \left(
  \begin{pmatrix}
    0 & 0 & 4 \\
    0 & -\sqrt{10} & 0 \\
    4 & 0 & 0 \\
  \end{pmatrix}
  H_u^2 -
  \begin{pmatrix}
    0 & \sqrt{10} & 0 \\
    \sqrt{10} & 0 & 0 \\
    0 & 0 & 2 \\
  \end{pmatrix}
  H_u^3 +
  \begin{pmatrix}
    4 & 0 & 0 \\
    0 & 0 & -2 \\
    0 & -2 & 0 \\
  \end{pmatrix}
  H_u^4
  \right) \notag \\
  &+ \sum_m \sqrt{2}\alpha_{u2}^{4_m}{\bf Y}_{4_m}^3(\omega) \left(
  -\begin{pmatrix}
    0 & 0 & 2 \\
    0 & \sqrt{10} & 0 \\
    2 & 0 & 0 \\
  \end{pmatrix}
  H_u^2 +
  \begin{pmatrix}
    0 & -\sqrt{10} & 0 \\
    -\sqrt{10} & 0 & 0 \\
    0 & 0 & 1 \\
  \end{pmatrix}
  H_u^3 +
  \begin{pmatrix}
    -2 & 0 & 0 \\
    0 & 0 & 1 \\
    0 & 1 & 0 \\
  \end{pmatrix}
  H_u^4 \right) \notag \\
  &+ \sum_n \sqrt{3}\alpha_{u1}^{5_n}
  \begin{pmatrix}
    0 & 0 & {\bf Y}_{5_n}^3(\omega) \\
    0 & \sqrt{2}{\bf Y}_{5_n}^2(\omega) & 0 \\
    {\bf Y}_{5_n}^3(\omega) & 0 & 0 \\
  \end{pmatrix} H_u^1 \notag \\
  &+ \sum_n 2\sqrt{3}\alpha_{u2}^{5_n} \left(
  \begin{pmatrix}
    0 & 0 & 0 \\
    0 & -\sqrt{2}{\bf Y}_{5_n}^2(\omega) & 0 \\
    {\bf Y}_{5_n}^3(\omega) & 0 & 0 \\
  \end{pmatrix}
  H_u^2 
  +
  \begin{pmatrix}
    0 & \sqrt{2}{\bf Y}_{5_n}^2(\omega) & 0 \\
    0 & 0 & 0 \\
    0 & 0 & -{\bf Y}_{5_n}^3(\omega) \\
  \end{pmatrix}
  H_u^3
  \right. \notag \\
  &\hspace{2.5 cm} + \left.\left.
  \begin{pmatrix}
    -{\bf Y}_{5_n}^3(\omega) & 0 & 0 \\
    0 & 0 & {\bf Y}_{5_n}^3(\omega) \\
    0 & 0 & 0 \\
  \end{pmatrix}
  H_u^4 \right)\right]
  \begin{pmatrix}
    u \\ c \\ t \\
  \end{pmatrix}.
\end{align}
To find NNI forms, here we assume $\langle H_u^3\rangle = 0$.
Then the up-type quark mass matrix is obtained as
\begin{align}
  M_u &= 
  \sum_k \alpha_{u1}^{3_k}{\bf Y}_{3_k}^2(\omega) 
  \begin{pmatrix}
    0 & 0 & -1 \\
    0 & 0 & 0 \\
    1 & 0 & 0 \\
  \end{pmatrix}
  \langle H_u^1\rangle \notag \\
  &+
  \sum_k \alpha_{u2}^{3_k}{\bf Y}_{3_k}^2(\omega) \left(
  \begin{pmatrix}
    0 & 0 & -4 \\
    0 & 0 & 0 \\
    -2 & 0 & 0 \\
  \end{pmatrix}
  \langle H_u^2\rangle +
  \begin{pmatrix}
    3 & 0 & 0 \\
    0 & 0 & -1 \\
    0 & 1 & 0 \\
  \end{pmatrix}
  \langle H_u^4\rangle \right) \notag \\
  &+\sum_\ell \frac{\alpha_{u2}^{3'_\ell}}{\sqrt{3}}{\bf Y}_{3'_\ell}^2(\omega)
  \left(
  \begin{pmatrix}
    0 & 0 & 4 \\
    0 & -\sqrt{10} & 0 \\
    4 & 0 & 0 \\
  \end{pmatrix}
  \langle H_u^2\rangle +
  \begin{pmatrix}
    4 & 0 & 0 \\
    0 & 0 & -2 \\
    0 & -2 & 0 \\
  \end{pmatrix}
  \langle H_u^4\rangle
  \right) \notag \\
  &+ \sum_m \sqrt{2}\alpha_{u2}^{4_m}{\bf Y}_{4_m}^3(\omega) \left(
  -\begin{pmatrix}
    0 & 0 & 2 \\
    0 & \sqrt{10} & 0 \\
    2 & 0 & 0 \\
  \end{pmatrix}
  \langle H_u^2\rangle +
  \begin{pmatrix}
    -2 & 0 & 0 \\
    0 & 0 & 1 \\
    0 & 1 & 0 \\
  \end{pmatrix}
  \langle H_u^4\rangle \right) \notag \\
  &+ \sum_n \sqrt{3}\alpha_{u1}^{5_n}
  \begin{pmatrix}
    0 & 0 & {\bf Y}_{5_n}^3(\omega) \\
    0 & \sqrt{2}{\bf Y}_{5_n}^2(\omega) & 0 \\
    {\bf Y}_{5_n}^3(\omega) & 0 & 0 \\
  \end{pmatrix} \langle H_u^1\rangle \notag \\
  &+ \sum_n 2\sqrt{3}\alpha_{u2}^{5_n} \left(
  \begin{pmatrix}
    0 & 0 & 0 \\
    0 & -\sqrt{2}{\bf Y}_{5_n}^2(\omega) & 0 \\
    {\bf Y}_{5_n}^3(\omega) & 0 & 0 \\
  \end{pmatrix}
  \langle H_u^2\rangle + 
  \begin{pmatrix}
    -{\bf Y}_{5_n}^3(\omega) & 0 & 0 \\
    0 & 0 & {\bf Y}_{5_n}^3(\omega) \\
    0 & 0 & 0 \\
  \end{pmatrix}
  \langle H_u^4\rangle \right).
\end{align}
This mass matrix has six kinds of independent parameters, $\alpha_{u1}^{3_k}$, $\alpha_{u2}^{3_k}$, $\alpha_{u2}^{3'_\ell}$, $\alpha_{u2}^{4_m}$, $\alpha_{u1}^{5_n}$ and $\alpha_{u2}^{5_n}$.
By taking appropriate liner combinations of them, it is possible to realize the following form matrix,
\begin{align}
  M_u =
  \begin{pmatrix}
    0 & 0 & \beta^1 \\
    0 & \gamma^1 & 0 \\
    \alpha^1 & 0 & 0 \\
  \end{pmatrix}
  +
  \begin{pmatrix}
    \alpha^2 & 0 & 0 \\
    0 & 0 & \gamma^2 \\
    0 & \beta^2 & 0 \\
  \end{pmatrix},
\end{align}
where $\alpha^{1,2}$, $\beta^{1,2}$ and $\gamma^{1,2}$ are liner functions of the above six parameters.
Therefore we can realize NNI forms in this $A_5$ model with four pairs of Higgs fields.

\section{Summary}

The NNI form is a desirable base  to derive the Fritzsch-type quark mass matrix with specific texture zeros
although the NNI form   is  a general form of quark mass matrices.
 We have studied the flavor models  of quarks systematically  to realize the NNI form of mass matrices explicitly.
The NNI form of quark mass matrices are derived 
 in modular flavor symmetric models at the fixed point $\tau = \omega$.
We  have presented models that the NNI forms of the quark mass matrices are simply realized  at the fixed point 
 $\tau=\omega$ in the $A_4$ modular flavor symmetry  by taking account  multi-Higgs fields.
 Those are also simple examples  that   the CP is violated  even at $\tau=\omega$ in the 
 case of the  finite modular symmetry with multi-Higgs fields.
 We show  that such  texture zero structure originates from  the  $ST$ charge of the residual symmetry
 $Z_3$ of $SL(2,Z)$.
The NNI form can be realized at the fixed point $\tau = \omega$ in $A_4$ and $S_4$ modular flavor models with two pairs of Higgs fields,
when we assign properly modular weights to Yukawa couplings and $A_4$ and $S_4$ representations to 
three generations of quarks.
We need four pairs of Higgs fields to realize the NNI form in $A_5$ modular flavor models.

Thus, the modular flavor models with multi-Higgs fields at the fixed point $\tau = \omega$
leads to successful  quark mass matrices.
We can extend our analysis to the lepton sector.
Extension to the charged lepton mass matrix and Dirac neutrino mass matrix is straightforward, 
and we obtain the same results.
Similarly, we can study the right-handed neutrino mass matrices.
For the right-handed neutrino sector, the symmetric assignments are possible in Table \ref{tab:AssignmentsRep}, 
i.e. cases I, IV, and VI, and parameters in the mass matrix should be symmetric.
Since the Higgs VEVs do not appear in the right-handed neutrino mass matrix, 
we replace the Higgs VEV of the trivial singlet by a constant with setting the Higgs VEVs with other representation to be zero.
Case VI has a limited form of the mass matrix with one free parameter, while case I has a generic $(3\times 3)$ symmetric 
mass matrix. 
It is interesting to study more phenomenological aspects on the flavor physics based on our systematical analysis 
at the fixed point $\tau = \omega$ in the near future.

\vspace{1.5 cm}
\noindent
{\large\bf Acknowledgement}\\

This work was supported by  JSPS KAKENHI Grant Numbers JP22J10172 (SK), and JP20J20388 (HU).

\appendix
\section*{Appendix}

\section{Group theoretical aspects}

Here, we give a review on group theoretical aspects of 
$A_4$, $S_4$, and $A_5$.
 
\subsection{$A_4$}
\label{app:A4group}

The generators of $A_4$ are denoted by $S$ and $T$, and 
they satisfy the following algebraic relations:
\begin{align}
  S^2 = (ST)^3 = T^3 = 1.
\end{align}
In $A_4$ group, there are four irreducible representations, three singlet $1$, $1''$ and $1'$ and one triplet $3$.
Each irreducible representation is given by
\begin{align}
  &1\quad \rho(S)=1, ~\rho(T)=1, \\
  &1'' \quad \rho(S)=1,~\rho(T)=\omega, \\
  &1' \quad \rho(S)=1, ~\rho(T)=\omega^2, \\
  &3 \quad 
  \rho(S) = \frac{1}{3}
  \begin{pmatrix}
    -1 & 2 & 2 \\
    2 & -1 & 2 \\
    2 & 2 & -1 \\
  \end{pmatrix},~
  \rho(T) =
  \begin{pmatrix}
    1 & 0 & 0 \\
    0 & \omega & 0 \\
    0 & 0 & \omega^2 \\
  \end{pmatrix},
\end{align}
in the $T$-diagonal basis.
In the  $ST$-diagonal basis, they have the following $ST$-eigenvalues.
\begin{align}
  &1\quad \rho(ST)=1, \\
  &1'' \quad \rho(ST)=\omega, \\
  &1' \quad \rho(ST)=\omega^2, \\
  &3 \quad 
  \rho(ST) =
  \begin{pmatrix}
    1 & 0 & 0 \\
    0 & \omega & 0 \\
    0 & 0 & \omega^2 \\
  \end{pmatrix}.
\end{align}
Their multiplication rules are shown in Table \ref{tab:MultiRuleinA4}.
\begin{table}[H]
\begin{center}
\renewcommand{\arraystretch}{1}
\begin{tabular}{c|c|c} \hline
  Tensor product  & $T$-diagonal basis & $ST$-diagonal basis \\ \hline
  $1'' \otimes 1'' = 1'$ & \multirow{3}{*}{$a^1b^1$} & \multirow{3}{*}{$a^1b^1$} \\
  $~~~~~~~~~~1' \otimes 1' = 1''$ ~~$(a^1 b^1)$ & & \\
  $1'' \otimes 1' = 1$ & & \\ \hline
  \multirow{2}{*}{$1'' \otimes 3 = 3$ ~~$(a^1 b^i)$} & \multirow{2}{*}{$\left(\begin{smallmatrix} a^1b^3\\ a^1b^1\\ a^1b^2\\\end{smallmatrix}\right)$} & \multirow{2}{*}{$\omega^2\left(\begin{smallmatrix} a^1b^3\\ a^1b^1\\ a^1b^2\\\end{smallmatrix}\right)$} \\
  & & \\ \hline
  \multirow{2}{*}{$1' \otimes 3 = 3$~~$(a^1 b^i)$} & \multirow{2}{*}{$\left(\begin{smallmatrix} a^1b^2\\ a^1b^3\\ a^1b^1\\ \end{smallmatrix}\right)$} & \multirow{2}{*}{$\omega\left(\begin{smallmatrix} a^1b^2\\ a^1b^3\\ a^1b^1 \\ \end{smallmatrix}\right)$} \\
  & & \\ \hline
  \multirow{5}{*}{$3\otimes 3=1\oplus 1'' \oplus 1' \oplus 3 \oplus 3$} & $\begin{smallmatrix}(a^1b^1+a^2b^3+a^3b^2)\end{smallmatrix}$ & $\begin{smallmatrix}\omega(a^1b^1+a^2b^3+a^3b^2) \\\end{smallmatrix}$ \\
  & $\oplus\begin{smallmatrix}(a^1b^2+a^2b^1+a^3b^3)\end{smallmatrix}$ & $\oplus\begin{smallmatrix}\omega^2(a^1b^2+a^2b^1+a^3b^3) \\\end{smallmatrix}$ \\
  & $\oplus\begin{smallmatrix}(a^1b^3+a^2b^2+a^3b^1)\end{smallmatrix}$ & $\oplus\begin{smallmatrix}(a^1b^3+a^2b^2+a^3b^1) \\\end{smallmatrix}$ \\
 \multirow{2}{*}{$(a^ib^j)$}  & \multirow{2}{*}{$\oplus\frac{1}{3}\left(\begin{smallmatrix} 2a^1b^1-a^2b^3-a^3b^2\\ -a^1b^2-a^2b^1+2a^3b^3\\ -a^1b^3+2a^2b^2-a^3b^1\\ \end{smallmatrix}\right)$} & \multirow{2}{*}{$\oplus\frac{\omega^2}{3}\left(\begin{smallmatrix} 
  2a^1b^1-a^2b^3-a^3b^2 \\
  -a^1b^2-a^2b^1+2a^3b^3 \\
  -a^1b^3+2a^2b^2-a^3b^1 \\ 
\end{smallmatrix}\right)$} \\
  & & \\
  & \multirow{2}{*}{$\oplus\frac{1}{2}\left(\begin{smallmatrix} a^2b^3-a^3b^2\\ a^1b^2-a^2b^1\\ -a^1b^3+a^3b^1\\ \end{smallmatrix}\right)$} & \multirow{2}{*}{$\oplus\frac{\omega^2}{2}
  \left(\begin{smallmatrix}
  a^2b^3-a^3b^2 \\
  a^1b^2-a^2b^1 \\
  -a^1b^3+a^3b^1 \\ 
  \end{smallmatrix}\right)$} \\
  & & \\ \hline
\end{tabular}
\end{center}
\caption{Multiplication rule in irreducible representations of $A_4$.
The second column shows decompositions  in the $T$-diagonal basis and the third column shows ones in the  $ST$-basis.}
\label{tab:MultiRuleinA4}
\end{table}

\subsection{$S_4$}
\label{app:S4group}

The generators $S$ and $T$ of $S_4$ group satisfy the following algebraic relations:
\begin{align}
  S^2 = (ST)^3 = T^4 = 1.
\end{align}
In $S_4$ group, there are five irreducible representations, two singlet $1$ and $1'$, one doublet $2$ and two triplets $3$ and $3'$.
Each irreducible representation is given by
\begin{align}
  &1\quad \rho(S)=1, ~\rho(T)=1, \\
  &1' \quad \rho(S)=-1, ~\rho(T)=-1, \\
  &2 \quad 
  \rho(S) = \frac{1}{2}
  \begin{pmatrix}
    -1 & \sqrt{3} \\
    \sqrt{3} & 1 \\
  \end{pmatrix},~
  \rho(T) =
  \begin{pmatrix}
    1 & 0 \\
    0 & -1 \\
  \end{pmatrix}, \\
  &3 \quad 
  \rho(S) = -\frac{1}{2}
  \begin{pmatrix}
    0 & \sqrt{2} & \sqrt{2} \\
    \sqrt{2} & -1 & 1 \\
    \sqrt{2} & 1 & -1 \\
  \end{pmatrix},~
  \rho(T) =
  \begin{pmatrix}
    -1 & 0 & 0 \\
    0 & -i & 0 \\
    0 & 0 & i \\
  \end{pmatrix}, \\
  &3' \quad 
  \rho(S) = \frac{1}{2}
  \begin{pmatrix}
    0 & \sqrt{2} & \sqrt{2} \\
    \sqrt{2} & -1 & 1 \\
    \sqrt{2} & 1 & -1 \\
  \end{pmatrix},~
  \rho(T) =
  \begin{pmatrix}
    1 & 0 & 0 \\
    0 & i & 0 \\
    0 & 0 & -i \\
  \end{pmatrix}, 
\end{align}
in the $T$-diagonal basis.
In the  $ST$-diagonal basis, they have the following $ST$-eigenvalues.
\begin{align}
  &1\quad \rho(ST)=1, \\
  &1' \quad \rho(ST)=1, \\
  &2 \quad 
  \rho(ST) = 
  \begin{pmatrix}
    \omega & 0 \\
    0 & \omega^2 \\
  \end{pmatrix}, \\
  &3 \quad 
  \rho(ST) = 
  \begin{pmatrix}
    1 & 0 & 0 \\
    0 & \omega & 0 \\
    0 & 0 & \omega^2 \\
  \end{pmatrix}, \\
  &3' \quad 
  \rho(ST) = 
  \begin{pmatrix}
    1 & 0 & 0 \\
    0 & \omega & 0 \\
    0 & 0 & \omega^2 \\
  \end{pmatrix},.
\end{align}
Their multiplication rules are shown  in Table \ref{tab:MultiRuleinS4}.
\begin{table}[H]
\begin{center}
\renewcommand{\arraystretch}{1}
\begin{tabular}{c|c|c} \hline
  Tensor product  & $T$-diagonal basis & $ST$-diagonal basis \\ \hline
  $1' \otimes 1' = 1$ ~~$(a^1 b^1)$& $a^1b^1$ & $a^1b^1$ \\ \hline
  $1' \otimes 2 = 2$ ~~$(a^1 b^i)$& $\left(\begin{smallmatrix} a^1b^2 \\ -a^1b^1 \\\end{smallmatrix}\right)$ & $-i\left(\begin{smallmatrix} a^1b^1 \\ -a^1b^2 \\\end{smallmatrix}\right)$ \\ \hline
  $1' \otimes 3 = 3'$ & \multirow{2}{*}{$\left(\begin{smallmatrix} a^1b^1 \\ a^1b^2 \\ a^1b^3 \\ \end{smallmatrix}\right)$} & \multirow{2}{*}{$\left(\begin{smallmatrix} a^1b^1 \\ a^1b^2 \\ a^1b^3 \\ \end{smallmatrix}\right)$} \\
  ~~~~~~~~~~$1' \otimes 3' = 3$~~$(a^1 b^i)$ & \\ \hline
  \multirow{3}{*}{$2 \otimes 2 = 1 \oplus 1' \oplus 2$} & $\frac{1}{\sqrt{2}}(\begin{smallmatrix} a^1b^1+a^2b^2 \end{smallmatrix})$ & $-\frac{1}{\sqrt{2}}(\begin{smallmatrix} a^1b^2+a^2b^1 \end{smallmatrix})$ \\
  & $\oplus \frac{1}{\sqrt{2}}(\begin{smallmatrix}a^1b^2-a^2b^1\end{smallmatrix})$ & $\oplus -\frac{i}{\sqrt{2}}(\begin{smallmatrix}a^1b^2-a^2b^1\end{smallmatrix})$ \\
\multirow{1}{*}{$(a^i b^j)$}  & \multirow{2}{*}{$\oplus \frac{1}{\sqrt{2}}\left(\begin{smallmatrix} a^1b^2+a^2b^1 \\ a^1b^1-a^2b^2 \\ \end{smallmatrix}\right)$} & \multirow{2}{*}{$\oplus -\frac{i}{\sqrt{2}}\left(\begin{smallmatrix} a^2b^2 \\ a^1b^1 \\ \end{smallmatrix}\right)$} \\
  & \\ \hline
  \multirow{3}{*}{$2 \otimes 3 = 3 \oplus 3'$} & \multirow{2}{*}{$\left(\begin{smallmatrix} a^1b^1 \\ (\sqrt{3}/2)a^2b^3-(1/2)a^1b^2 \\ (\sqrt{3}/2)a^2b^2-(1/2)a^1b^3 \end{smallmatrix}\right)$} & \multirow{2}{*}{$\frac{1}{\sqrt{2}}\left(\begin{smallmatrix} -a^1b^3+a^2b^2 \\ -a^1b^1+a^2b^3 \\ -a^1b^2+a^2b^1 \end{smallmatrix}\right)$} \\
  & \\
 \multirow{1}{*}{$(a^i b^j)$} & \multirow{2}{*}{$\oplus\left(\begin{smallmatrix} -a^2b^1 \\ (\sqrt{3}/2)a^1b^3+(1/2)a^2b^2 \\ (\sqrt{3}/2)a^1b^2+(1/2)a^2b^3 \end{smallmatrix}\right)$} & \multirow{2}{*}{$\oplus-\frac{i}{\sqrt{2}}\left(\begin{smallmatrix} a^1b^3+a^2b^2 \\ a^1b^1+a^2b^3 \\ a^1b^2+a^2b^1 \end{smallmatrix}\right)$} \\
  &  \\ \hline
  \multirow{3}{*}{$2 \otimes 3' = 3 \oplus 3'$} & \multirow{2}{*}{$\left(\begin{smallmatrix} -a^2b^1 \\ (\sqrt{3}/2)a^1b^3+(1/2)a^2b^2 \\ (\sqrt{3}/2)a^1b^2+(1/2)a^2b^3 \end{smallmatrix}\right)$} & \multirow{2}{*}{$-\frac{i}{\sqrt{2}}\left(\begin{smallmatrix} a^1b^3+a^2b^2 \\ a^1b^1+a^2b^3 \\ a^1b^2+a^2b^1 \end{smallmatrix}\right)$} \\ 
  & \\
\multirow{1}{*}{$(a^i b^j)$}  & \multirow{2}{*}{$\oplus\left(\begin{smallmatrix} a^1b^1 \\ (\sqrt{3}/2)a^2b^3-(1/2)a^1b^2 \\ (\sqrt{3}/2)a^2b^2-(1/2)a^1b^3 \end{smallmatrix}\right)$} & \multirow{2}{*}{$\oplus\frac{1}{\sqrt{2}}\left(\begin{smallmatrix} -a^1b^3+a^2b^2 \\ -a^1b^1+a^2b^3 \\ -a^1b^2+a^2b^1 \end{smallmatrix}\right)$} \\
  & \\ \hline
  & \multirow{3}{*}{$\begin{matrix}\begin{smallmatrix}\frac{1}{\sqrt{3}}(a^1b^1+a^2b^3+a^3b^2)\end{smallmatrix}\\ \oplus\left(\begin{smallmatrix}(2a^1b^1-a^2b^3-a^3b^2)/\sqrt{6}\\ (a^2b^2+a^3b^3)/\sqrt{2}\\\end{smallmatrix}\right)\end{matrix}$} & \multirow{3}{*}{$\begin{matrix}\begin{smallmatrix}\frac{1}{\sqrt{3}}(a^1b^1+a^2b^3+a^3b^2)\end{smallmatrix}\\ \oplus \frac{1}{\sqrt{3}}\left(\begin{smallmatrix}-a^1b^2-a^2b^1-a^3b^3 \\ a^1b^3+a^2b^2+a^3b^1 \\\end{smallmatrix}\right)\end{matrix}$} \\
  & \\
  $3 \otimes 3 = 1 \oplus 2 \oplus 3 \oplus 3'$ & \\
  $3' \otimes 3' = 1 \oplus 2 \oplus 3 \oplus 3'$ & \multirow{2}{*}{$\oplus \frac{1}{\sqrt{2}}\left(\begin{smallmatrix}a^3b^3-a^2b^2\\ a^1b^3+a^3b^1\\ -a^1b^2-a^2b^1\\\end{smallmatrix}\right)$} & \multirow{2}{*}{$\oplus -i\left(\begin{smallmatrix}(-2a^1b^1+a^2b^3+a^3b^2)/\sqrt{6} \\ (a^1b^2+a^2b^1-2a^3b^3)/\sqrt{6} \\ (a^1b^3-2a^2b^2+a^3b^1)/\sqrt{6} \\\end{smallmatrix}\right)$} \\
  $(a^ib^j)$ & \\
  & \multirow{2}{*}{$\oplus \frac{1}{\sqrt{2}}\left(\begin{smallmatrix}a^3b^2-a^2b^3\\ a^2b^1-a^1b^2\\ a^1b^3-a^3b^1\\\end{smallmatrix}\right)$} & \multirow{2}{*}{$\oplus \frac{1}{\sqrt{2}}\left(\begin{smallmatrix}a^2b^3-a^3b^2 \\ a^1b^2-a^2b^1\\ -a^1b^3+a^3b^1\\\end{smallmatrix}\right)$} \\
  & \\ \hline
  \multirow{5}{*}{$3 \otimes 3' = 1' \oplus 2 \oplus 3 \oplus 3'$} & \multirow{3}{*}{$\begin{matrix}\begin{smallmatrix}\frac{1}{\sqrt{3}}(a^1b^1+a^2b^3+a^3b^2)\end{smallmatrix}\\ \oplus\left(\begin{smallmatrix} (a^2b^2+a^3b^3)/\sqrt{2}\\ (-2a^1b^1+a^2b^3+a^3b^2)/\sqrt{6}\\\end{smallmatrix}\right)\end{matrix}$} & \multirow{3}{*}{$\begin{matrix}\begin{smallmatrix}\frac{1}{\sqrt{3}}(a^1b^1+a^2b^3+a^3b^2)\end{smallmatrix}\\ \oplus\frac{i}{\sqrt{3}}\left(\begin{smallmatrix} a^1b^2+a^2b^1+a^3b^3 \\ a^1b^3+a^2b^2+a^3b^1 \\\end{smallmatrix}\right)\end{matrix}$} \\
  & \\
  & \\
\multirow{2}{*}{$(a^i b^j)$}   & \multirow{2}{*}{$\oplus \frac{1}{\sqrt{2}}\left(\begin{smallmatrix}a^3b^2-a^2b^3\\ a^2b^1-a^1b^2\\ a^1b^3-a^3b^1\\\end{smallmatrix}\right)$} & \multirow{2}{*}{$\oplus \frac{1}{\sqrt{2}}\left(\begin{smallmatrix}a^2b^3-a^3b^2 \\ a^1b^2-a^2b^1\\ -a^1b^3+a^3b^1\\\end{smallmatrix}\right)$} \\
  & \\
 & \multirow{2}{*}{$\oplus \frac{1}{\sqrt{2}}\left(\begin{smallmatrix}a^3b^3-a^2b^2\\ a^1b^3+a^3b^1\\ -a^1b^2-a^2b^1\\\end{smallmatrix}\right)$} & \multirow{2}{*}{$\oplus -i\left(\begin{smallmatrix}(-2a^1b^1+a^2b^3+a^3b^2)/\sqrt{6} \\ (a^1b^2+a^2b^1-2a^3b^3)/\sqrt{6} \\ (a^1b^3-2a^2b^2+a^3b^1)/\sqrt{6} \\\end{smallmatrix}\right)$} \\
  & \\ \hline
\end{tabular}
\end{center}
\caption{Multiplication rule in irreducible representations of $S_4$.
The second column shows decompositions  in the $T$-diagonal basis and the third column shows ones in the  $ST$-diagonal basis.}
\label{tab:MultiRuleinS4}
\end{table}

\subsection{$A_5$}
\label{app:A5group}

The generators of $A_5$ are denoted by $S$ and $T$, and they satisfy the following algebraic relations:
\begin{align}
  S^2 = (ST)^3 = T^5 = 1.
\end{align}
In $A_5$ group, there are five irreducible representations, one singlet $1$, two triplets $3$ and $3'$, one four-dimensional representation $4$ and one five-dimensional representation $5$.
Each irreducible representation is given by
\begin{align}
  &1\quad \rho(S)=1, ~\rho(T)=1, \\
  &3 \quad 
  \rho(S) = \frac{1}{\sqrt{5}}
  \begin{pmatrix}
    1 & -\sqrt{2} & -\sqrt{2} \\
    -\sqrt{2} & -\phi_g & \phi_g-1 \\
    -\sqrt{2} & \phi_g-1 &-\phi_g \\
  \end{pmatrix},~
  \rho(T) =
  \begin{pmatrix}
    1 & 0 & 0 \\
    0 & \omega_5 & 0 \\
    0 & 0 & \omega_5^4 \\
  \end{pmatrix}, \\
  &3' \quad 
  \rho(S) = \frac{1}{\sqrt{5}}
  \begin{pmatrix}
    -1 & \sqrt{2} & \sqrt{2} \\
    \sqrt{2} & 1-\phi_g & \phi_g \\
    \sqrt{2} & \phi_g & 1-\phi_g \\
  \end{pmatrix},~
  \rho(T) =
  \begin{pmatrix}
    1 & 0 & 0 \\
    0 & \omega_5^2 & 0 \\
    0 & 0 & \omega_5^3 \\
  \end{pmatrix}, \\
  &4 \quad
  \rho(S) = \frac{1}{\sqrt{5}}
  \begin{pmatrix}
    1 & \phi_g-1 & \phi_g & -1 \\
    \phi_g-1 & -1 & 1 & \phi_g \\
    \phi_g & 1 & -1 & \phi_g-1 \\
    -1 & \phi_g & \phi_g-1 & 1 \\
  \end{pmatrix},~
  \rho(T) =
  \begin{pmatrix}
    \omega_5 & 0 & 0 & 0 \\
    0 & \omega_5^2 & 0 & 0 \\
    0 & 0 & \omega_5^3 & 0 \\
    0 & 0 & 0 & \omega_5^4 \\
  \end{pmatrix}, \\
  &5 \quad
  \rho(S) = \frac{1}{5}
  \begin{pmatrix}
    -1 & \sqrt{6} & \sqrt{6} & \sqrt{6} & \sqrt{6} \\
    \sqrt{6} & (\phi_g-1)^2 & -2\phi_g & 2(\phi_g-1) & \phi_g^2 \\
    \sqrt{6} & -2\phi_g & \phi_g^2 & (\phi_g-1)^2 & 2(\phi_g-1) \\
    \sqrt{6} & 2(\phi_g-1) & (\phi_g-1)^2 & \phi_g^2 & -2\phi_g \\
    \sqrt{6} & \phi_g^2 & 2(\phi_g-1) & -2\phi_g & (\phi_g-1)^2 \\
  \end{pmatrix},~
  \rho(T) =
  \begin{pmatrix}
    1 & 0 & 0 & 0 & 0 \\
    0 & \omega_5 & 0 & 0 & 0 \\
    0 & 0 & \omega_5^2 & 0 & 0 \\
    0 & 0 & 0 & \omega_5^3 & 0 \\
    0 & 0 & 0 & 0 & \omega_5^4 \\
  \end{pmatrix},
\end{align}
in the $T$-diagonal basis.
Here we have used $\omega_5 = e^{2\pi i/5}$ and $\phi_g = (1+\sqrt{5})/2$.
In the $ST$-diagonal basis, they have the following $ST$-eigenvalues.
\begin{align}
  &1\quad \rho(ST)=1, \\
  &3 \quad 
  \rho(ST) = 
  \begin{pmatrix}
    1 & 0 & 0 \\
    0 & \omega & 0 \\
    0 & 0 & \omega^2 \\
  \end{pmatrix}, \\
  &3' \quad 
  \rho(ST) = 
  \begin{pmatrix}
    1 & 0 & 0 \\
    0 & \omega & 0 \\
    0 & 0 & \omega^2 \\
  \end{pmatrix}, \\
  &4 \quad 
  \rho(ST) = 
  \begin{pmatrix}
    1 & 0 & 0 & 0 \\
    0 & 1 & 0 & 0 \\
    0 & 0 & \omega & 0 \\
    0 & 0 & 0 & \omega^2 \\
  \end{pmatrix}, \\
  &5\quad
  \rho(ST) =
  \begin{pmatrix}
    1 & 0 & 0 & 0 & 0 \\
    0 & \omega & 0 & 0 & 0 \\
    0 & 0 & \omega & 0 & 0 \\
    0 & 0 & 0 & \omega^2 & 0 \\
    0 & 0 & 0 & 0 & \omega^2 \\
  \end{pmatrix}.
\end{align}
Their multiplication rules are shown in Table \ref{tab:MultiRuleinA5}.
\begin{table}[H]
\begin{center}
\caption{Multiplication rule in irreducible representations of $A_5$.
The first column shows tensor product decompositions in the $T$-diagonal basis and the second column shows ones in the $ST$-diagonal basis.}
\label{tab:MultiRuleinA5}
\renewcommand{\arraystretch}{1}
\begin{tabular}{c|c} \hline
  \multicolumn{2}{c}{$3 \otimes 3 =1\oplus 3\oplus 5 \quad (a^ib^j)$} \\ \hline
  $T$-diagonal basis & $ST$-diagonal basis \\ \hline
  $\begin{matrix} \begin{smallmatrix} a^1b^1+a^2b^3+a^3b^2 \end{smallmatrix} \\ \oplus  \left(\begin{smallmatrix} a^2b^3-a^3b^2 \\ a^1b^2-a^2b^1 \\ a^3b^1-a^1b^3 \\ \end{smallmatrix}\right) \\ \oplus  \left(\begin{smallmatrix} 2a^1b^1-a^2b^3-a^3b^2 \\ -\sqrt{3}a^1b^2-\sqrt{3}a^2b^1 \\ \sqrt{6}a^2b^2 \\ \sqrt{6}a^3b^3 \\ -\sqrt{3}a^1b^3-\sqrt{3}a^3b^1 \\ \end{smallmatrix}\right) \\ \end{matrix}$
  & $\begin{matrix} \begin{smallmatrix}
  a^1b^1+a^2b^3+a^3b^2 \\
\end{smallmatrix} \\ \oplus \left(\begin{smallmatrix}
  a^2b^3-a^3b^2 \\
  a^1b^2-a^2b^1 \\
  -a^1b^3+a^3b^1 \\
\end{smallmatrix}\right) \\ \oplus 
\left( \begin{smallmatrix}
  2a^1b^1-a^2b^3-a^3b^2 \\
  \sqrt{6}a^3b^3 \\
  \sqrt{3}a^1b^2+\sqrt{3}a^2b^1 \\
  \sqrt{6}a^2b^2 \\
  \sqrt{3}a^1b^3+\sqrt{3}a^3b^1 \\
\end{smallmatrix} \right)
\end{matrix}$ \\ \hline
  \multicolumn{2}{c}{$3' \otimes 3' = 1\oplus 3'\oplus 5 \quad (a^ib^j)$} \\ \hline
  $T$-diagonal basis & $ST$-diagonal basis \\ \hline
  $\begin{matrix} \begin{smallmatrix} a^1b^1+a^2b^3+a^3b^2 \end{smallmatrix} \\ \oplus  \left(\begin{smallmatrix} a^2b^3-a^3b^2 \\ a^1b^2-a^2b^1 \\ a^3b^1-a^1b^3 \end{smallmatrix}\right) \\ \oplus \left(\begin{smallmatrix} 2a^1b^1-a^2b^3-a^3b^2 \\ \sqrt{6}a^3b^3 \\ -\sqrt{3}a^1b^2-\sqrt{3}a^2b^1 \\ -\sqrt{3}a^1b^3-\sqrt{3}a^3b^1 \\ \sqrt{6}a^2b^2 \\ \end{smallmatrix}\right) \\ \end{matrix}$
  & $\begin{matrix} 
  \begin{smallmatrix}
  a^1b^1+a^2b^3+a^3b^2 \\
\end{smallmatrix} \\ \oplus 
  \left(
  \begin{smallmatrix}
  a^2b^3-a^3b^2 \\
  a^1b^2-a^2b^1 \\
  -a^1b^3+a^3b^1 \\
\end{smallmatrix}
  \right) \\ \oplus 
  \left(
  \begin{smallmatrix}
  2a^1b^1-a^2b^3-a^3b^2 \\
  \sqrt{6}a^3b^3 \\
  \sqrt{3}a^1b^2+\sqrt{3}a^2b^1 \\
  \sqrt{6}a^2b^2 \\
  \sqrt{3}a^1b^3+\sqrt{3}a^3b^1 \\
\end{smallmatrix}
  \right)
  \end{matrix}$ \\ \hline
  \multicolumn{2}{c}{$3 \otimes 3' = 4\oplus 5 \quad (a^ib^j)$} \\ \hline
  $T$-diagonal basis & $ST$-diagonal basis \\ \hline
  $\begin{matrix} \left(\begin{smallmatrix} \sqrt{2}a^2b^1+a^3b^2 \\ -\sqrt{2}a^1b^2-a^3b^3 \\ -\sqrt{2}a^1b^3-a^2b^2 \\ \sqrt{2}a^3b^1+a^2b^3 \\ \end{smallmatrix}\right) \\ \oplus \left(\begin{smallmatrix} \sqrt{3}a^1b^1 \\ a^2b^1-\sqrt{2}a^3b^2 \\ a^1b^2-\sqrt{2}a^3b^3 \\ a^1b^3-\sqrt{2}a^2b^2 \\ a^3b^1-\sqrt{2}a^2b^3 \\ \end{smallmatrix}\right) \\ \end{matrix}$
  & $\begin{matrix}
  \frac{1}{\sqrt{3}}\left( \begin{smallmatrix}
  (3/\sqrt{2})(a^2b^3-a^3b^2) \\
  (1/\sqrt{2})(-4a^1b^1+a^2b^3+a^3b^2) \\
  \sqrt{2}a^1b^2-\sqrt{2}a^2b^1+\sqrt{5}a^3b^3 \\
  \sqrt{2}a^1b^3+\sqrt{5}a^2b^2-\sqrt{2}a^3b^1 \\
\end{smallmatrix} \right) \\ \oplus
  \frac{1}{3}\left(
  \begin{smallmatrix}
  \sqrt{3}a^1b^1+2\sqrt{3}a^2b^3+2\sqrt{3}a^3b^2 \\
  -\sqrt{5}a^1b^2-2\sqrt{5}a^2b^1-\sqrt{2}a^3b^3 \\
  4a^1b^2-a^2b^1-\sqrt{10}a^3b^3 \\
  -\sqrt{5}a^1b^3-\sqrt{2}a^2b^2-2\sqrt{5}a^3b^1 \\
  4a^1b^3-\sqrt{10}a^2b^2-a^3b^1 \\
\end{smallmatrix}
  \right)
  \end{matrix}$\\ \hline
  \multicolumn{2}{c}{$3 \otimes 4 = 3'\oplus 4\oplus 5 \quad (a^ib^j)$} \\ \hline
  $T$-diagonal basis & $ST$-diagonal basis \\ \hline
  $\begin{matrix} \left(\begin{smallmatrix} -\sqrt{2}a^2b^4-\sqrt{2}a^3b^1 \\ \sqrt{2}a^1b^2-a^2b^1+a^3b^3 \\ \sqrt{2}a^1b^3+a^2b^2-a^3b^4 \\ \end{smallmatrix} \right) \\ \oplus \left(\begin{smallmatrix} a^1b^1-\sqrt{2}a^3b^2 \\ -a^1b^2-\sqrt{2}a^2b^1 \\ a^1b^3+\sqrt{2}a^3b^4 \\ -a^1b^4+\sqrt{2}a^2b^3 \\  \end{smallmatrix}\right) \\ \oplus \left(\begin{smallmatrix} \sqrt{6}a^2b^4-\sqrt{6}a^3b^1 \\ 2\sqrt{2}a^1b^1 + 2a^3b^2 \\ -\sqrt{2}a^1b^2+a^2b^1+3a^3b^3 \\ \sqrt{2}a^1b^3-3a^2b^2-a^3b^4 \\ -2\sqrt{2}a^1b^4-2a^2b^3 \\  \end{smallmatrix}\right) \\ \end{matrix}$
  & $\begin{matrix}
  \frac{1}{\sqrt{6}}\left(
  \begin{smallmatrix}
  4a^1b^2+2a^2b^4+2a^3b^3 \\
  -2a^1b^3+3a^2b^1-a^2b^2-\sqrt{10}a^3b^4 \\
  -2a^1b^4-\sqrt{10}a^2b^3-3a^3b^1-a^3b^2 \\
\end{smallmatrix}
  \right) \\ \oplus
  \left(
  \begin{smallmatrix}
  a^1b^2-a^2b^4-a^3b^3 \\
  a^1b^1+a^2b^4-a^3b^3 \\
  -a^1b^3-a^2b^1-a^2b^2 \\
  a^1b^4-a^3b^1+a^3b^2 \\
\end{smallmatrix}
  \right) \\ \oplus
  \frac{1}{\sqrt{6}}\left(
  \begin{smallmatrix}
  4\sqrt{3}a^1b^1-2\sqrt{3}a^2b^4+2\sqrt{3}a^3b^3 \\
  2\sqrt{5}a^1b^3+\sqrt{5}a^2b^1-3\sqrt{5}a^2b^2+\sqrt{2}a^3b^4 \\
  4a^1b^3-4a^2b^1-2\sqrt{10}a^3b^4 \\
  -2\sqrt{5}a^1b^4-\sqrt{2}a^2b^3+\sqrt{5}a^3b^1+3\sqrt{5}a^3b^2 \\
  -4a^1b^4+2\sqrt{10}a^2b^3-4a^3b^1 \\
\end{smallmatrix}
  \right)
  \end{matrix}$ \\ \hline 
\end{tabular}
\end{center}
\end{table}
\begin{table}[H]
\begin{center}
\renewcommand{\arraystretch}{1}
\begin{tabular}{c|c} \hline
  \multicolumn{2}{c}{$3' \otimes 4 = 3\oplus 4\oplus 5 \quad (a^ib^j)$} \\ \hline 
  $T$-diagonal basis & $ST$-diagonal basis \\ \hline
  $\begin{matrix} \left(\begin{smallmatrix} -\sqrt{2}a^2b^3-\sqrt{2}a^3b^2 \\ \sqrt{2}a^1b^1+a^2b^4-a^3b^3 \\ \sqrt{2}a^1b^4-a^2b^2+a^3b^1 \\ \end{smallmatrix}\right) \\ \oplus \left(\begin{smallmatrix} a^1b^1+\sqrt{2}a^3b^3 \\ a^1b^2-\sqrt{2}a^3b^4 \\ -a^1b^3+\sqrt{2}a^2b^1 \\ -a^1b^4-\sqrt{2}a^2b^2 \\ \end{smallmatrix}\right) \\ \oplus \left(\begin{smallmatrix} \sqrt{6}a^2b^3-\sqrt{6}a^3b^2 \\ \sqrt{2}a^1b^1-3a^2b^4 - a^3b^3 \\ 2\sqrt{2}a^1b^2+2a^3b^4 \\ -2\sqrt{2}a^1b^3-2a^2b^1 \\ -\sqrt{2}a^1b^4+a^2b^2+3a^3b^1 \\ \end{smallmatrix}\right) \\ \end{matrix}$
  & $\begin{matrix}
  \frac{1}{\sqrt{6}}\left(
  \begin{smallmatrix}
  -4a^1b^2+2a^2b^4+2a^3b^3 \\
  -2a^1b^3+3a^2b^1+a^2b^2+\sqrt{10}a^3b^4 \\
  -2a^1b^4+\sqrt{10}a^2b^3-3a^3b^1+a^3b^2 \\
\end{smallmatrix}
  \right) \\ \oplus
  \left(
  \begin{smallmatrix}
  -a^1b^2-a^2b^4-a^3b^3 \\
  -a^1b^1-a^2b^4+a^3b^3 \\
  -a^1b^3-a^2b^1+a^2b^2 \\
  a^1b^4-a^3b^1-a^3b^2 \\
\end{smallmatrix}
  \right) \\ \oplus
  \frac{1}{\sqrt{6}}\left(
  \begin{smallmatrix}
  4\sqrt{3}a^1b^1-2\sqrt{3}a^2b^4+2\sqrt{3}a^3b^3 \\
  2\sqrt{5}a^2b^1+2\sqrt{5}a^2b^2-4\sqrt{2}a^3b^4 \\
  -6a^1b^3+a^2b^1-5a^2b^2-\sqrt{10}a^3b^4 \\
  4\sqrt{2}a^2b^3+2\sqrt{5}a^3b^1-2\sqrt{5}a^3b^2 \\
  6a^1b^4+\sqrt{10}a^2b^3+a^3b^1+5a^3b^2 \\
\end{smallmatrix}
  \right)
  \end{matrix}$ \\ \hline
  \multicolumn{2}{c}{$3 \otimes 5 = 3\oplus 3'\oplus 4\oplus 5 \quad (a^ib^j)$} \\ \hline
  $T$-diagonal basis & $ST$-diagonal basis \\ \hline
  $\begin{matrix} \left(\begin{smallmatrix} -2a^1b^1+\sqrt{3}a^2b^5+\sqrt{3}a^3b^2 \\ \sqrt{3}a^1b^2+a^2b^1-\sqrt{6}a^3b^3 \\ \sqrt{3}a^1b^5-\sqrt{6}a^2b^4+a^3b^1 \\ \end{smallmatrix}\right) \\ \oplus 
  \left(\begin{smallmatrix} \sqrt{3}a^1b^1+a^2b^5+a^3b^2 \\ a^1b^3-\sqrt{2}a^2b^2-\sqrt{2}a^3b^4 \\ a^1b^4-\sqrt{2}a^2b^3-\sqrt{2}a^3b^5 \\ \end{smallmatrix}\right) \\ \oplus 
  \left(\begin{smallmatrix} 2\sqrt{2}a^1b^2-\sqrt{6}a^2b^1+a^3b^3 \\ -\sqrt{2}a^1b^3+2a^2b^2-3a^3b^4 \\ \sqrt{2}a^1b^4+3a^2b^3-2a^3b^5 \\ -2\sqrt{2}a^1b^5-a^2b^4+\sqrt{6}a^3b^1 \\ \end{smallmatrix}\right) \\ \oplus 
  \left(\begin{smallmatrix} \sqrt{3}a^2b^5-\sqrt{3}a^3b^2 \\ -a^1b^2-\sqrt{3}a^2b^1-\sqrt{2}a^3b^3 \\ -2a^1b^3-\sqrt{2}a^2b^2 \\ 2a^1b^4+\sqrt{2}a^3b^5 \\ a^1b^5+\sqrt{2}a^2b^4+\sqrt{3}a^3b^1 \\ \end{smallmatrix}\right) \\ \end{matrix}$
  & $\begin{matrix}
  \left(
  \begin{smallmatrix}
  -2a^1b^1-\sqrt{3}a^2b^5-\sqrt{3}a^3b^3 \\
  -\sqrt{3}a^1b^3+a^2b^1-\sqrt{6}a^3b^4 \\
  -\sqrt{3}a^1b^5-\sqrt{6}a^2b^2+a^3b^1 \\
\end{smallmatrix}
  \right) \\ \oplus 
  \frac{1}{3}\left(
  \begin{smallmatrix}
  \sqrt{3}a^1b^1-2\sqrt{5}a^2b^4-a^2b^5-2\sqrt{5}a^3b^2-a^3b^3 \\
  -\sqrt{5}a^1b^2+4a^1b^3+2\sqrt{3}a^2b^1-\sqrt{2}a^3b^4-\sqrt{10}a^3b^5 \\
  -\sqrt{5}a^1b^4+4a^1b^5-\sqrt{2}a^2b^2-\sqrt{10}a^2b^3+2\sqrt{3}a^3b^1 \\
\end{smallmatrix}
  \right) \\ \oplus 
  \frac{1}{\sqrt{6}}\left(
  \begin{smallmatrix}
  4\sqrt{3}a^1b^1+\sqrt{5}a^2b^4-4a^2b^5+\sqrt{5}a^3b^2-4a^3b^3 \\
  3\sqrt{5}a^2b^4-3\sqrt{5}a^3b^2 \\
  2\sqrt{5}a^1b^2+4a^1b^3+2\sqrt{3}a^2b^1-\sqrt{2}a^3b^4+2\sqrt{10}a^3b^5 \\
  -2\sqrt{5}a^1b^4-4a^1b^5+\sqrt{2}a^2b^2-2\sqrt{10}a^2b^3-2\sqrt{3}a^3b^1 \\
\end{smallmatrix}
  \right) \\ \oplus 
  \left(
  \begin{smallmatrix}
  -\sqrt{3}a^2b^5+\sqrt{3}a^3b^3 \\
  2a^1b^2-\sqrt{2}a^3b^5 \\
  -a^1b^3+\sqrt{3}a^2b^1+\sqrt{2}a^3b^4 \\
  -2a^1b^4+\sqrt{2}a^2b^3 \\
  a^1b^5-\sqrt{2}a^2b^2-\sqrt{3}a^3b^1 \\
\end{smallmatrix}
  \right)
  \end{matrix}$ \\ \hline
  \multicolumn{2}{c}{$3' \otimes 5 = \oplus 3'\oplus 4\oplus 5 \quad (a^ib^j)$} \\ \hline
  $T$-diagonal basis & $ST$-diagonal basis \\ \hline
  $\begin{matrix} \left(\begin{smallmatrix} \sqrt{3}a^1b^1+a^2b^4+a^3b^3 \\ a^1b^2-\sqrt{2}a^2b^5-\sqrt{2}a^3b^4 \\ a^1b^5-\sqrt{2}a^2b^3-\sqrt{2}a^3b^2 \\ \end{smallmatrix}\right) \\ \oplus 
  \left(\begin{smallmatrix} -2a^1b^1+\sqrt{3}a^2b^4+\sqrt{3}a^3b^3 \\ \sqrt{3}a^1b^3+a^2b^1-\sqrt{6}a^3b^5 \\ \sqrt{3}a^1b^4-\sqrt{6}a^2b^2+a^3b^1 \\ \end{smallmatrix}\right) \\ \oplus 
  \left(\begin{smallmatrix} \sqrt{2}a^1b^2+3a^2b^5-2a^3b^4 \\ 2\sqrt{2}a^1b^3-\sqrt{6}a^2b^1+a^3b^5 \\ -2\sqrt{2}a^1b^4-a^2b^2+\sqrt{6}a^3b^1 \\ -\sqrt{2}a^1b^5+2a^2b^3-3a^3b^2 \\ \end{smallmatrix}\right) \\ \oplus 
  \left(\begin{smallmatrix} \sqrt{3}a^2b^4-\sqrt{3}a^3b^3 \\ 2a^1b^2+\sqrt{2}a^3b^4 \\ -a^1b^3-\sqrt{3}a^2b^1-\sqrt{2}a^3b^5 \\ a^1b^4+\sqrt{2}a^2b^2+\sqrt{3}a^3b^1 \\ -2a^1b^5-\sqrt{2}a^2b^3 \\ \end{smallmatrix}\right) \\\end{matrix}$
  & $\begin{matrix}
  \frac{1}{3}\left(
  \begin{smallmatrix}
  \sqrt{3}a^1b^1-\sqrt{5}a^2b^4+4a^2b^5-\sqrt{5}a^3b^2+4a^3b^3 \\
  -2\sqrt{5}a^1b^2-a^1b^3+2\sqrt{3}a^2b^1-\sqrt{2}a^3b^4-\sqrt{10}a^3b^5 \\
  -2\sqrt{5}a^1b^4-a^1b^5-\sqrt{2}a^2b^2-\sqrt{10}a^2b^3+2\sqrt{3}a^3b^1 \\
\end{smallmatrix}
  \right) \\ \oplus 
  \frac{1}{\sqrt{6}}\left(
  \begin{smallmatrix}
  2\sqrt{6}a^1b^1-\sqrt{10}a^2b^4-2\sqrt{2}a^2b^5-\sqrt{10}a^3b^2-2\sqrt{2}a^3b^3 \\
  -\sqrt{10}a^1b^2-2\sqrt{2}a^1b^3-\sqrt{6}a^2b^1-4a^3b^4+2\sqrt{5}a^3b^5 \\
  -\sqrt{10}a^1b^4-2\sqrt{2}a^1b^5-4a^2b^2+2\sqrt{5}a^2b^3-\sqrt{6}a^3b^1 \\
\end{smallmatrix}
  \right) \\ \oplus 
  \frac{1}{\sqrt{6}}\left(
  \begin{smallmatrix}
  4\sqrt{3}a^1b^1+2\sqrt{5}a^2b^4+a^2b^5+2\sqrt{5}a^3b^2+a^3b^3 \\
  -2\sqrt{5}a^2b^4+5a^2b^5+2\sqrt{5}a^3b^2-5a^3b^3 \\
  -a^1b^3+2\sqrt{3}a^2b^1+4\sqrt{2}a^3b^4+\sqrt{10}a^3b^5 \\
  a^1b^5-4\sqrt{2}a^2b^2-\sqrt{10}a^2b^3-2\sqrt{3}a^3b^1 \\
\end{smallmatrix}
  \right) \\ \oplus 
  \frac{1}{3}\left(
  \begin{smallmatrix}
  \sqrt{15}a^2b^4+2\sqrt{3}a^2b^5-\sqrt{15}a^3b^2-2\sqrt{3}a^3b^3 \\
  a^1b^2-2\sqrt{5}a^1b^3-\sqrt{15}a^2b^1-3\sqrt{2}a^3b^5 \\
  -2\sqrt{5}a^1b^2+2a^1b^3-2\sqrt{3}a^2b^1+3\sqrt{2}a^3b^4 \\
  -a^1b^4+2\sqrt{5}a^1b^5+3\sqrt{2}a^2b^3+\sqrt{15}a^3b^1 \\
  2\sqrt{5}a^1b^4-2a^1b^5-3\sqrt{2}a^2b^2+2\sqrt{3}a^3b^1 \\
\end{smallmatrix}
  \right) \\
  \end{matrix}$ \\ \hline
\end{tabular}
\end{center}
\end{table}
\begin{table}[H]
\begin{center}
\renewcommand{\arraystretch}{1}
\begin{tabular}{c|c} \hline
    \multicolumn{2}{c}{$4 \otimes 4 = 1\oplus 3\oplus 3'\oplus 4\oplus 5 \quad (a^ib^j)$} \\ \hline
  $T$-diagonal basis & $ST$-diagonal basis \\ \hline
  $\begin{matrix} \begin{smallmatrix} a^1b^4+a^2b^3+a^3b^2+a^4b^1 \end{smallmatrix} \\ \oplus
  \left(\begin{smallmatrix} -a^1b^4+a^2b^3-a^3b^2+a^4b^1 \\ \sqrt{2}a^2b^4-\sqrt{2}a^4b^2 \\ \sqrt{2}a^1b^3-\sqrt{2}a^3b^1 \\ \end{smallmatrix}\right) \\ \oplus
  \left(\begin{smallmatrix} a^1b^4+a^2b^3-a^3b^2-a^4b^1 \\ \sqrt{2}a^3b^4-\sqrt{2}a^4b^3 \\ \sqrt{2}a^1b^2-\sqrt{2}a^2b^1 \\ \end{smallmatrix}\right) \\ \oplus 
  \left(\begin{smallmatrix} a^2b^4+a^3b^3+a^4b^2 \\ a^1b^1+a^3b^4+a^4b^3 \\ a^1b^2+a^2b^1 + a^4b^4 \\  a^1b^3+a^2b^2+a^3b^1 \\ \end{smallmatrix}\right) \\ \oplus 
  \left(\begin{smallmatrix} \sqrt{3}(a^1b^4-a^2b^3-a^3b^2+a^4b^1) \\ \sqrt{2}(-a^2b^4+2a^3b^3-a^4b^2) \\ \sqrt{2}(-2a^1b^1+a^3b^4+a^4b^3) \\ \sqrt{2}(a^1b^2+a^2b^1-2a^4b^4) \\ \sqrt{2}(-a^1b^3+2a^2b^2-a^3b^1) \\  \end{smallmatrix}\right) \\ \end{matrix}$
  & $\begin{matrix}
  \begin{smallmatrix}
  -a^1b^1+a^2b^2+a^3b^4+a^4b^3 \\
\end{smallmatrix} \\ \oplus 
  \left(
  \begin{smallmatrix}
  -a^1b^2+a^2b^1+a^3b^4-a^4b^3 \\
  a^1b^3-a^2b^3-a^3b^1+a^3b^2 \\
  a^1b^4+a^2b^4-a^4b^1-a^4b^2 \\
\end{smallmatrix}
  \right) \\ \oplus 
  \left(
  \begin{smallmatrix}
  -a^1b^2+a^2b^1-a^3b^4+a^4b^3 \\
  -a^1b^3-a^2b^3+a^3b^1+a^3b^2 \\
  -a^1b^4+a^2b^4+a^4b^1-a^4b^2 \\
\end{smallmatrix}
  \right) \\ \oplus 
  \frac{1}{\sqrt{6}}\left(
  \begin{smallmatrix}
  -3a^1b^2-3a^2b^1 \\
  3a^1b^1-a^2b^2+2a^3b^4+2a^4b^3 \\
  2a^2b^3+2a^3b^2-\sqrt{10}a^4b^4 \\
  2a^2b^4-\sqrt{10}a^3b^3+2a^4b^2 \\
\end{smallmatrix}
  \right) \\ \oplus 
  \frac{1}{3}\left(
  \begin{smallmatrix}
  -3\sqrt{3}a^1b^1-5\sqrt{3}a^2b^2+\sqrt{3}a^3b^4+\sqrt{3}a^4b^3 \\
  3\sqrt{5}a^1b^3-\sqrt{5}a^2b^3+3\sqrt{5}a^3b^1-\sqrt{5}a^3b^2-2\sqrt{2}a^4b^4 \\
  -3a^1b^3-5a^2b^3-3a^3b^1-5a^3b^2-2\sqrt{10}a^4b^4 \\
  -3\sqrt{5}a^1b^4-\sqrt{5}a^2b^4-2\sqrt{2}a^3b^3-3\sqrt{5}a^4b^1-\sqrt{5}a^4b^2 \\
  3a^1b^4-5a^2b^4-2\sqrt{10}a^3b^3+3a^4b^1-5a^4b^2 \\
\end{smallmatrix}
  \right)
  \end{matrix}$ \\ \hline
  \multicolumn{2}{c}{$4 \otimes 5 = 3\oplus 3'\oplus 4\oplus 5\oplus 5 \quad (a^ib^j)$} \\ \hline
  $T$-diagonal basis & $ST$-diagonal basis \\ \hline
  $\begin{matrix}
\left(\begin{smallmatrix} 2\sqrt{2}a^1b^5-\sqrt{2}a^2b^4+\sqrt{2}a^3b^3-2\sqrt{2}a^4b^2 \\ -\sqrt{6}a^1b^1+2a^2b^5+3a^3b^4-a^4b^3 \\ a^1b^4-3a^2b^3-2a^3b^2+\sqrt{6}a^4b^1 \\ \end{smallmatrix}\right) \\ \oplus \\
\left(\begin{smallmatrix} \sqrt{2}a^1b^5+2\sqrt{2}a^2b^4-2\sqrt{2}a^3b^3-\sqrt{2}a^4b^2 \\ 3a^1b^2-\sqrt{6}a^2b^1-a^3b^5+2a^4b^4 \\ -2a^1b^3+a^2b^2+\sqrt{6}a^3b^1-3a^4b^5 \\  \end{smallmatrix}\right) \\ \oplus \\
\left(\begin{smallmatrix} \sqrt{3}a^1b^1-\sqrt{2}a^2b^5+\sqrt{2}a^3b^4-2\sqrt{2}a^4b^3 \\ -\sqrt{2}a^1b^2-\sqrt{3}a^2b^1+2\sqrt{2}a^3b^5+\sqrt{2}a^4b^4 \\ \sqrt{2}a^1b^3+2\sqrt{2}a^2b^2-\sqrt{3}a^3b^1-\sqrt{2}a^4b^5 \\ -2\sqrt{2}a^1b^4+\sqrt{2}a^2b^3-\sqrt{2}a^3b^2+\sqrt{3}a^4b^1 \\ \end{smallmatrix}\right) \\ \oplus \\
\left(\begin{smallmatrix} \sqrt{2}a^1b^5-\sqrt{2}a^2b^4-\sqrt{2}a^3b^3+\sqrt{2}a^4b^2 \\ -\sqrt{2}a^1b^1-\sqrt{3}a^3b^4-\sqrt{3}a^4b^3 \\ \sqrt{3}a^1b^2+\sqrt{2}a^2b^1+\sqrt{3}a^3b^5 \\ \sqrt{3}a^2b^2+\sqrt{2}a^3b^1+\sqrt{3}a^4b^5 \\ -\sqrt{3}a^1b^4-\sqrt{3}a^2b^3-\sqrt{2}a^4b^1 \\  \end{smallmatrix}\right) \\ \oplus \\
\left(\begin{smallmatrix} 2a^1b^5+4a^2b^4+4a^3b^3+2a^4b^2 \\ 4a^1b^1+2\sqrt{6}a^2b^5 \\ -\sqrt{6}a^1b^2+2a^2b^1-\sqrt{6}a^3b^5+2\sqrt{6}a^4b^4 \\ 2\sqrt{6}a^1b^3-\sqrt{6}a^2b^2+2a^3b^1-\sqrt{6}a^4b^5 \\ 2\sqrt{6}a^3b^2+4a^4b^1 \\ \end{smallmatrix}\right) \\
\end{matrix}$
  & $\begin{matrix}  
  \frac{1}{\sqrt{6}}\left(
  \begin{smallmatrix}
  4\sqrt{3}a^1b^1+2\sqrt{5}a^3b^4+4a^3b^5-2\sqrt{5}a^4b^2-4a^4b^3 \\
  \sqrt{5}a^1b^2-4a^1b^3+3\sqrt{5}a^2b^2+2\sqrt{3}a^3b^1+\sqrt{2}a^4b^4-2\sqrt{10}a^4b^5 \\
  \sqrt{5}a^1b^4-4a^1b^5-3\sqrt{5}a^2b^4-\sqrt{2}a^3b^2+2\sqrt{10
}a^3b^3-2\sqrt{3}a^4b^1 \\
\end{smallmatrix}
  \right) \\ \oplus \\
  \frac{1}{\sqrt{6}}\left(
  \begin{smallmatrix}
  4\sqrt{3}a^1b^1-6a^3b^5+6a^4b^3 \\
  2\sqrt{5}a^1b^2+a^1b^3-2\sqrt{5}a^2b^2+5a^2b^3+2\sqrt{3}a^3b^1-4\sqrt{2}a^4b^4-\sqrt{10}a^4b^5 \\
  2\sqrt{5}a^1b^4+a^1b^5+2\sqrt{5}a^2b^4-5a^2b^5+4\sqrt{2}a^3b^2+\sqrt{10}a^3b^3-2\sqrt{3}a^4b^1 \\
\end{smallmatrix}
  \right) \\ \oplus \\
  \frac{1}{3}\left(
  \begin{smallmatrix}
  3\sqrt{3}a^1b^1-3\sqrt{5}a^3b^4+3a^3b^5+3\sqrt{5}a^4b^2-3a^4b^3 \\
  -5\sqrt{3}a^2b^1-\sqrt{5}a^3b^4-5a^3b^5-\sqrt{5}a^4b^2-5a^4b^3 \\
  -3\sqrt{5}a^1b^2+3a^1b^3-\sqrt{5}a^2b^2-5a^2b^3+\sqrt{3}a^3b^1-2\sqrt{2}a^4b^4-2\sqrt{10}a^4b^5 \\
  (3\sqrt{5}a^1b^4-3a^1b^5-\sqrt{5}a^2b^4-5a^2b^5-2\sqrt{2}a^3b^2-2\sqrt{10}a^3b^3+\sqrt{3}a^4b^1 \\
\end{smallmatrix}
  \right) \\ \oplus \\
  \left(
  \begin{smallmatrix}
  \begin{smallmatrix}
  -\frac{2\sqrt{5}}{\sqrt{6}}a^3b^4+\frac{\sqrt{6}}{3}a^3b^5-\frac{2\sqrt{5}}{\sqrt{6}}a^4b^2+\frac{\sqrt{6}}{3}a^4b^3 \\
  \frac{2\sqrt{2}}{3}a^1b^2+\frac{\sqrt{10}}{6}a^1b^3-\frac{5}{\sqrt{10}}a^2b^3+\frac{2\sqrt{5}}{\sqrt{6}}a^3b^1+a^4b^5 \\
  \frac{\sqrt{10}}{6}a^1b^2+\frac{4\sqrt{2}}{3}a^1b^3+\frac{5}{\sqrt{10}}a^2b^2-\frac{\sqrt{6}}{3}a^3b^1-a^4b^4 \\
  -\frac{2\sqrt{2}}{3}a^1b^4-\frac{\sqrt{10}}{6}a^1b^5-\frac{5}{\sqrt{10}}a^2b^5+a^3b^3+\frac{2\sqrt{5}}{\sqrt{6}}a^4b^1 \\
  -\frac{\sqrt{10}}{6}a^1b^4-\frac{4\sqrt{2}}{3}a^1b^5+\frac{5}{\sqrt{10}}a^2b^4-a^3b^2-\frac{\sqrt{6}}{3}a^4b^1 \\
\end{smallmatrix}
\end{smallmatrix}
  \right) \\ \oplus \\
  \left(
  \begin{smallmatrix}
  4a^2b^1-\frac{6}{\sqrt{3}}a^3b^5-\frac{6}{\sqrt{3}}a^4b^3 \\
  -\frac{2}{3}a^1b^2-\frac{5\sqrt{5}}{3}a^1b^3+\frac{2}{3}a^2b^2-\frac{\sqrt{5}}{3}a^2b^3-\frac{2\sqrt{5}}{\sqrt{3}}a^3b^1-\frac{4\sqrt{10}}{3}a^4b^4+\frac{2}{3\sqrt{2}}a^4b^5 \\
  \frac{4\sqrt{5}}{3}a^1b^2-\frac{4}{3}a^1b^3-\frac{4\sqrt{5}}{3}a^2b^2-\frac{8}{3}a^2b^3-\frac{4}{\sqrt{3}}a^3b^1+\frac{4\sqrt{2}}{3}a^4b^4-\frac{2\sqrt{10}}{3}a^4b^5 \\
  \frac{2}{3}a^1b^4+\frac{5\sqrt{5}}{3}a^1b^5+\frac{2}{3}a^2b^4-\frac{\sqrt{5}}{3}a^2b^5-\frac{4\sqrt{10}}{3}a^3b^2+\frac{2}{3\sqrt{2}}a^3b^3-\frac{2\sqrt{5}}{\sqrt{3}}a^4b^1 \\
  -\frac{4\sqrt{5}}{3}a^1b^4+\frac{4}{3}a^1b^5-\frac{4\sqrt{5}}{3}a^2b^4-\frac{8}{3}a^2b^5+\frac{4\sqrt{2}}{3}a^3b^2-\frac{2\sqrt{10}}{3}a^3b^3-\frac{4}{\sqrt{3}}a^4b^1 \\
\end{smallmatrix}
  \right) \\
  \end{matrix}$ \\ \hline
\end{tabular}
\end{center}
\end{table}
\begin{table}[H]
\begin{center}
\renewcommand{\arraystretch}{1}
\begin{tabular}{c|c} \hline
  \multicolumn{2}{c}{$5 \otimes 5 = 1\oplus 3\oplus 3'\oplus 4\oplus 4\oplus 5\oplus 5 \quad (a^ib^j)$} \\ \hline
  $T$-diagonal basis & $ST$-diagonal basis \\ \hline
  $\begin{matrix}\begin{smallmatrix} a^1b^1+a^2b^5+a^3b^4+a^4b^3+a^5b^2 \end{smallmatrix}\\ \oplus \\ 
  \left(\begin{smallmatrix} a^2b^5+2a^3b^4-2a^4b^3-a^5b^2\\ -\sqrt{3}a^1b^2+\sqrt{3}a^2b^1+\sqrt{2}a^3b^5-\sqrt{2}a^5b^3\\ \sqrt{3}a^1b^5+\sqrt{2}a^2b^4-\sqrt{2}a^4b^2-\sqrt{3}a^5b^1 \\\end{smallmatrix}\right)\\ \oplus \\ 
  \left(\begin{smallmatrix} 2a^2b^5-a^3b^4+a^4b^3-2a^5b^2\\ \sqrt{3}a^1b^3-\sqrt{3}a^3b^1+\sqrt{2}a^4b^5-\sqrt{2}a^5b^4\\ -\sqrt{3}a^1b^4+\sqrt{2}a^2b^3-\sqrt{2}a^3b^2+\sqrt{3}a^4b^1 \\\end{smallmatrix}\right) \\ \oplus \\ 
  \left(\begin{smallmatrix} \sqrt{3}(\sqrt{6}(a^1b^2+a^2b^1)-a^3b^5+4a^4b^4-a^5b^3)\\ \sqrt{3}(\sqrt{6}(a^1b^3+a^3b^1)+4a^2b^2-a^4b^5-a^5b^4)\\ \sqrt{3}(\sqrt{6}(a^1b^4+a^4b^1)-a^2b^3-a^3b^2+4a^5b^5)\\ \sqrt{3}(\sqrt{6}(a^1b^5+a^5b^1)-a^2b^4+4a^3b^3-a^4b^2)\\ \end{smallmatrix}\right) \\ \oplus  \\ 
  \left(\begin{smallmatrix} \sqrt{2}a^1b^2-\sqrt{2}a^2b^1+\sqrt{3}a^3b^5-\sqrt{3}a^5b^3\\ -\sqrt{2}a^1b^3+\sqrt{2}a^3b^1+\sqrt{3}a^4b^5-\sqrt{3}a^5b^4\\ -\sqrt{2}a^1b^4-\sqrt{3}a^2b^3+\sqrt{3}a^3b^2+\sqrt{2}a^4b^1\\ \sqrt{2}a^1b^5-\sqrt{3}a^2b^4+\sqrt{3}a^4b^2-\sqrt{2}a^5b^1\\ \end{smallmatrix}\right) \\ \oplus \\ 
  \left(\begin{smallmatrix} 2a^1b^1+a^2b^5-2a^3b^4-2a^4b^3+a^5b^2\\ a^1b^2+a^2b^1+\sqrt{6}a^3b^5+\sqrt{6}a^5b^3\\ -2a^1b^3+\sqrt{6}a^2b^2-2a^3b^1\\ -2a^1b^4-2a^4b^1+\sqrt{6}a^5b^5\\ a^1b^5+\sqrt{6}a^2b^4+\sqrt{6}a^4b^2+a^5b^1\\\end{smallmatrix}\right) \\ \oplus \\ 
  \left(\begin{smallmatrix} 2a^1b^1-2a^2b^5+a^3b^4+a^4b^3-2a^5b^2\\ -2a^1b^2-2a^2b^1+\sqrt{6}a^4b^4\\ a^1b^3+a^3b^1+\sqrt{6}a^4b^5+\sqrt{6}a^5b^4\\ a^1b^4+\sqrt{6}a^2b^3+\sqrt{6}a^3b^2+a^4b^1\\ -2a^1b^5+\sqrt{6}a^3b^3-2a^5b^1\\ \end{smallmatrix}\right) \\ \end{matrix}$ 
  & $\begin{matrix}  
  \begin{smallmatrix}
  a^1b^1+a^2b^4+a^3b^5+a^4b^2+a^5b^3 \\
  \end{smallmatrix}
  \\ \oplus \\
  \left(
  \begin{smallmatrix}
  -2a^2b^4+a^3b^5+2a^4b^2-a^5b^3 \\
  \sqrt{3}a^1b^3-\sqrt{3}a^3b^1-\sqrt{2}a^4b^5+\sqrt{2}a^5b^4 \\
  -\sqrt{3}a^1b^5+\sqrt{2}a^2b^3-\sqrt{2}a^3b^2+\sqrt{3}a^5b^1 \\
\end{smallmatrix}
  \right) \\ \oplus \\
  \left(
  \begin{smallmatrix}
  \frac{1}{3}a^2b^4-\frac{2\sqrt{5}}{3}a^2b^5-\frac{2\sqrt{5}}{3}a^3b^4+\frac{2}{3}a^3b^5-\frac{1}{3}a^4b^2+\frac{2\sqrt{5}}{3}a^4b^3+\frac{2\sqrt{5}}{3}a^5b^2-\frac{2}{3}a^5b^3 \\
  \frac{\sqrt{5}}{\sqrt{3}}a^1b^2+\frac{2}{\sqrt{3}}a^1b^3-\frac{\sqrt{5}}{\sqrt{3}}a^2b^1-\frac{2}{\sqrt{3}}a^3b^1+\sqrt{2}a^4b^5-\sqrt{2}a^5b^4 \\
  -\frac{\sqrt{5}}{\sqrt{3}}a^1b^4-\frac{2}{\sqrt{3}}a^1b^5-\sqrt{2}a^2b^3+\sqrt{2}a^3b^2+\frac{\sqrt{5}}{\sqrt{3}}a^4b^1+\frac{2}{\sqrt{3}}a^5b^1 \\
\end{smallmatrix}
  \right) \\ \oplus \\
  \left(
  \begin{smallmatrix}
  \frac{3\sqrt{10}}{2}(-a^2b^5+a^3b^4+a^4b^3-a^5b^2) \\
  4\sqrt{2}a^1b^1+\frac{2\sqrt{2}}{3}(a^2b^4+a^4b^2)-\frac{5\sqrt{10}}{6}(a^2b^5+a^3b^4+a^4b^3+a^5b^2)-\frac{8\sqrt{2}}{3}(a^3b^5+a^5b^3) \\
  -\frac{2\sqrt{5}}{\sqrt{6}}(a^1b^2+a^2b^1)-\frac{5\sqrt{6}}{3}(a^1b^3+a^3b^1)-\frac{8\sqrt{5}}{3}a^4b^4+\frac{5}{3}(a^4b^5+a^5b^4)-\frac{4\sqrt{5}}{3}a^5b^5 \\
  -\frac{2\sqrt{5}}{\sqrt{6}}(a^1b^4+a^4b^1)-\frac{5\sqrt{6}}{3}(a^1b^5+a^5b^1)-\frac{8\sqrt{5}}{3}a^2b^2+\frac{5}{3}(a^2b^3+a^3b^2)-\frac{4\sqrt{5}}{3}a^3b^3 \\
\end{smallmatrix}
  \right) \\ \oplus \\
  \left(
  \begin{smallmatrix}
  \frac{2\sqrt{2}}{3}(a^2b^4-a^4b^2)+\frac{\sqrt{10}}{6}(a^2b^5-a^5b^2+a^3b^4-a^4b^3)+\frac{4\sqrt{2}}{3}(a^3b^5-a^5b^3) \\
  \frac{5}{\sqrt{10}}(-a^2b^5+a^5b^2+a^3b^4-a^4b^3) \\
  \frac{2\sqrt{5}}{\sqrt{6}}(-a^1b^2+a^2b^1)+\frac{\sqrt{6}}{3}(a^1b^3-a^3b^1)+a^4b^5-a^5b^4 \\
  \frac{2\sqrt{5}}{\sqrt{6}}(-a^1b^4+a^4b^1)+\frac{\sqrt{6}}{3}(a^1b^5-a^5b^1)+a^2b^3-a^3b^2 \\
\end{smallmatrix}
  \right) \\ \oplus \\
  \left(
  \begin{smallmatrix}
  2a^1b^1-2a^2b^4+a^3b^5-2a^4b^2+a^5b^3 \\
  -2a^1b^2-2a^2b^1+\sqrt{6}a^5b^5 \\
  a^1b^3+a^3b^1+\sqrt{6}a^4b^5+\sqrt{6}a^5b^4 \\
  -2a^1b^4+\sqrt{6}a^3b^3-2a^4b^1 \\
  a^1b^5+\sqrt{6}a^2b^3+\sqrt{6}a^3b^2+a^5b^1 \\
\end{smallmatrix}
  \right) \\ \oplus \\
  \left(
  \begin{smallmatrix}
  -2a^1b^1+\frac{1}{3}(a^2b^4+a^4b^2)-\frac{2\sqrt{5}}{3}(a^2b^5+a^5b^2+a^3b^4+a^4b^3)+\frac{2}{3}(a^3b^5+a^5b^3) \\
  \frac{1}{3}(a^1b^2+a^2b^1)-\frac{2\sqrt{5}}{3}(a^1b^3+a^3b^1)+\frac{10\sqrt{6}}{9}a^4b^4+\frac{\sqrt{30}}{9}(a^4b^5+a^5b^4)-\frac{8}{3\sqrt{6}}a^5b^5 \\
  -\frac{2\sqrt{5}}{3}(a^1b^2+a^2b^1)+\frac{2}{3}(a^1b^3+a^3b^1)+\frac{\sqrt{30}}{9}a^4b^4-\frac{8}{3\sqrt{6}}(a^4b^5+a^5b^4)-\frac{8\sqrt{5}}{3\sqrt{6}}a^5b^5 \\
  \frac{1}{3}(a^1b^4+a^4b^1)-\frac{2\sqrt{5}}{3}(a^1b^5+a^5b^1)+\frac{10\sqrt{6}}{9}a^2b^2+\frac{5\sqrt{6}}{9\sqrt{5}}(a^2b^3+a^3b^2)-\frac{8}{3\sqrt{6}}a^3b^3 \\
  -\frac{2\sqrt{5}}{3}(a^1b^4+a^4b^1)+\frac{2}{3}(a^1b^5+a^5b^1)+\frac{5\sqrt{6}}{9\sqrt{5}}a^2b^2-\frac{8}{3\sqrt{6}}(a^2b^3+a^3b^2)-\frac{8\sqrt{5}}{3\sqrt{6}}a^3b^3 \\
\end{smallmatrix}
  \right) \\
  \end{matrix}$ \\ \hline
\end{tabular}
\end{center}
\end{table}


\section{Modular forms of level $N=3$}
\label{app:ModularForms3}

Here, we give a brief review on the modular forms of $A_4$.
The modular forms of the weight 2, which correspond to the $A_4$ triplet, are written by
\begin{align}
  {\bf Y}^{\bf (2)}_3 =
  \begin{pmatrix}
    Y_1(\tau) \\ Y_2(\tau) \\ Y_3(\tau) \\
  \end{pmatrix}
\end{align}
where
\begin{align}
  &Y_1(\tau) = \frac{i}{2\pi} \left(\frac{\eta'(\tau/3)}{\eta(\tau/3)}+\frac{\eta'((\tau+1)/3)}{\eta((\tau+1)/3)}+\frac{\eta'((\tau+2)/3)}{\eta((\tau+2)/3)}-\frac{27\eta'(3\tau)}{\eta(3\tau)}\right), \\
  &Y_2(\tau) = \frac{-i}{\pi} \left(\frac{\eta'(\tau/3)}{\eta(\tau/3)}+\omega^2\frac{\eta'((\tau+1)/3)}{\eta((\tau+1)/3)}+\omega\frac{\eta'((\tau+2)/3)}{\eta((\tau+2)/3)}\right), \\
  &Y_3(\tau) = \frac{-i}{\pi} \left(\frac{\eta'(\tau/3)}{\eta(\tau/3)}+\omega\frac{\eta'((\tau+1)/3)}{\eta((\tau+1)/3)}+\omega^2\frac{\eta'((\tau+2)/3)}{\eta((\tau+2)/3)}\right).
\end{align}

The modular forms of higher weights can be obtained by products of the above modular forms as  
\begin{align}
  &{\bf Y}^{\bf (4)}_{1} = Y^2_1+2Y_2Y_3, \quad
  {\bf Y}^{\bf (4)}_{1'} = Y^2_3+2Y_1Y_2, \\
  &{\bf Y}^{\bf (4)}_{3}
  =
  \begin{pmatrix}
    Y^2_1-Y_2Y_3 \\ Y^2_3-Y_1Y_2 \\ Y^2_2-Y_1Y_3 \\
  \end{pmatrix}, \\
  &{\bf Y}^{\bf (6)}_{3,1} = (Y^2_1+2Y_2Y_3)
  \begin{pmatrix}
    Y_1 \\ Y_2 \\ Y_3 \\
  \end{pmatrix}, \quad
  {\bf Y}^{\bf (6)}_{3,2} = (Y^2_3+2Y_1Y_2)
  \begin{pmatrix}
    Y_3 \\ Y_1 \\ Y_2 \\
  \end{pmatrix}, \\
  &{\bf Y}^{\bf (6)}_1 = Y^3_1+Y^3_2+Y^3_3-3Y_1Y_2Y_3.
\end{align}


\end{document}